\documentclass[twocolumn,showpacs,floatfix,amsmath,amssymb,prb,superscriptaddress]{revtex4}
\usepackage[dvips]{graphicx}
\usepackage{latexsym}
\usepackage{graphicx}
\usepackage{times,psfrag,subfigure}
\usepackage{amsmath}
\usepackage{dcolumn}
\usepackage{latexsym,amsmath,amssymb,bm,euscript}
\def\etal{~\textit{et~al.}} 
\def\ra{\rangle} 
\def\la{\langle} 

\def\Hc{{\rm h.c.}}
\def\bfx{{\bf x}}
\def\barpsi{\overline{\psi}}

\begin{document}

\title{Criticality in quantum triangular antiferromagnets via
fermionized vortices}
\author{Jason Alicea}
\affiliation{Physics Department, University of California, Santa
Barbara, CA 93106}
\author{Olexei I. Motrunich}
\affiliation{Kavli Institute for Theoretical Physics, University of
California, Santa Barbara, CA 93106}
\author{Michael Hermele}
\affiliation{Physics Department, University of California, Santa
Barbara, CA 93106}
\author{Matthew P. A. Fisher}
\affiliation{Kavli Institute for Theoretical Physics, University of
California, Santa
Barbara, CA 93106}

\date{\today}

\begin{abstract}
We revisit two-dimensional frustrated quantum magnetism from a new 
perspective, with the aim of exploring new critical points and 
critical phases.  We study easy-plane triangular antiferromagnets 
using a dual vortex approach, \emph{fermionizing the vortices} with a 
Chern-Simons field.  Herein we develop this technique
for integer-spin systems which generically exhibit a simple paramagnetic phase as well as
magnetically-ordered phases with coplanar and collinear spin order.    
Within the fermionized-vortex approach, 
we derive a low-energy effective theory containing
Dirac fermions with two flavors minimally coupled 
to a $U(1)$ and a Chern-Simons gauge field.  
At criticality we argue that the Chern-Simons gauge field can be subsumed into
the $U(1)$ gauge field, and up to irrelevant interactions one arrives
at quantum electrodynamics in (2+1) dimensions (QED3).
Moreover, we conjecture that critical 
QED3 with full $SU(2)$ flavor symmetry describes the $O(4)$ multicritical 
point of the spin model where the paramagnet and two magnetically-ordered
phases merge.  The remarkable implication is that \emph{QED3 with flavor SU(2) symmetry
is dual to ordinary   
critical $\Phi^4$ field theory with $O(4)$ symmetry}.  This leads to 
a number of unexpected, verifiable predictions for QED3.
A connection of our fermionized-vortex approach 
with the dipole interpretation of the $\nu=1/2$ fractional quantum Hall state
is also demonstrated.
The approach introduced in this paper will be applied to
spin-1/2 systems in a forthcoming publication.
\end{abstract}

\maketitle

\section{Introduction}

Geometrically frustrated two-dimensional (2D) quantum spin systems 
provide promising settings for the realization of exotic quantum 
phenomena due to the frustration-enhanced role of quantum
fluctuations.  The quasi-2D spin-1/2 triangular antiferromagnet
$\text{Cs}_2\text{CuCl}_4$ and the organic Mott insulator
$\kappa\text{-(ET)}_2\text{Cu}_2(\text{CN})_3$ provide two
noteworthy examples that have generated much interest.  While 
$\text{Cs}_2\text{CuCl}_4$ exhibits ``cycloid'' magnetic order at low
temperatures,
evidence of a spin-liquid phase with deconfined spinon excitations 
was found at higher temperatures.\cite{Coldea,Coldea2}
Moreover, application of an in-plane magnetic field suppressed 
magnetic order down to the lowest temperatures achieved, which might
indicate a stabilization of the spin-liquid phase.\cite{Coldea}
In the organic material 
$\kappa\text{-(ET)}_2\text{Cu}_2(\text{CN})_3$,
NMR measurements show no signs of magnetic ordering, suggesting a 
spin-liquid phase as a candidate ground state.\cite{Kanoda}

Understanding critical spin-liquid states in 2D frustrated spin-1/2
systems is a central issue that has been the focus of significant 
theoretical attention in response to the above experiments.  Several
authors have attempted to access spin-liquid phases in these systems 
using slave-fermion and slave-boson mean field
approaches,\cite{Chung1,Chung2,Wen}
and recent work\cite{Isakov} has studied universal properties 
near the quantum phase transition between a spiral magnetic state 
and one such spin liquid with gapped bosonic spinons.
Spin-liquid states were also recently examined  using slave-particle
mean field approaches in the triangular lattice Hubbard model near 
the Mott transition, which is argued to be relevant for 
$\kappa\text{-(ET)}_2\text{Cu}_2(\text{CN})_3$.\cite{Lee,Motrunich}  

The purpose of this paper is to introduce a new dual vortex approach 
to frustrated quantum spin systems that allows one to gain insight 
into critical spin-liquid phases coming from the easy-plane regime.  
Our goal will be to illustrate how one can formulate and
analyze an effective low-energy dual theory describing criticality 
in frustrated quantum systems using \emph{fermionized vortices}.
As we will see, an advantage of this approach is that 
critical points and critical phases arise 
rather naturally from this perspective.

In the present paper we develop the fermionized-vortex 
approach by studying easy-plane \emph{integer-spin} triangular 
antiferromagnets, postponing an analysis of the more challenging 
spin-1/2 systems for a future publication.  
In the integer-spin case, a direct Euclidean formulation of the
quantum spin model is free of Berry phases, and therefore 
equivalent to a classical system of stacked triangular
antiferromagnets.  Thus, a fairly complete analysis is accessible 
in the direct spin language.  
Generically, such systems exhibit a paramagnetic phase, a
magnetically-ordered phase with collinear spin order, and the 
well-known coplanar $\sqrt{3}\times\sqrt{3}$ state, and much is known
about the intervening phase transitions.  

Using the direct analysis as a guide, we are led to a number of 
surprising results in the fermionized-vortex formulation, which we now 
summarize.  We conjecture that criticality in the spin model is 
described by critical quantum electrodynamics in (2+1) dimensions (QED3) with 
two flavors of two-component Dirac fermions minimally coupled to a 
noncompact gauge field.  In particular, we conjecture that this
critical QED3 theory describes the $O(4)$ multicritical point of the spin model
where the paramagnet merges with the coplanar and collinear ordered
phases.  In other words, we propose that critical QED3 with $SU(2)$ flavor
symmetry is dual to $O(4)$ critical $\Phi^4$ theory.  
Some remarkable predictions for the former theory follow from this 
conjecture, and can in principle be used to test its validity.  
Namely, we predict that single-monopole insertions in the 
QED3 theory, which add discrete units of $2\pi$ gauge flux, have the
same scaling dimension as the ordering field in the $O(4)$ theory.
In fact, the correspondence between the two theories can be stated
by observing that there are two distinct monopole insertions in the QED3, 
associating a complex field with each monopole, and grouping the two 
complex fields into an $O(4)$ vector.
Furthermore, fermionic bilinears $\barpsi\bm{\tau}\psi$, which are not
part of any conserved current, and double-strength monopoles in QED3 are
predicted to have \emph{identical} scaling dimensions, equal to that
of the quadratic anisotropy fields at the $O(4)$ fixed point.  
It would be quite remarkable if the scaling dimensions for these
seemingly unrelated operators indeed merge at the QED3 fixed point.

Given the length of this paper, it will be worthwhile to now provide 
an overview of the fermionized-vortex approach and describe the main
steps that lead to our results for the integer-spin triangular
antiferromagnet.  We will first illustrate how one obtains a low-energy
description of fermionized vortices, and then present a more 
detailed summary of the
integer-spin case.  We will also mention below an application of this 
approach in a different context, namely treating bosons at filling 
factor $\nu = 1$, which reproduces the so-called ``dipole picture''
of the compressible quantum Hall fluid. 

We start by highlighting the difficulties associated with frustration 
encountered in a conventional duality approach.  
Frustrated easy-plane spin models are equivalent to systems of bosons 
hopping in a magnetic field inducing frustration, and can be readily 
cast into the dual vortex representation.\cite{duality}  
A crucial feature of the resulting dual theory is that the vortices 
are at \emph{finite density} due to the underlying frustration.  
The finite vortex density together with the strong vortex interactions 
present a serious obstacle in treating the bosonic vortices, as it is not
clear how one can obtain a low-energy description of such degrees of
freedom that describes criticality in the spin system.  
However, we will demonstrate that
\emph{fermionizing} the vortices via Chern-Simons flux attachment 
allows one to successfully construct a low-energy dual vortex 
description in both the integer-spin \emph{and} spin-1/2 cases.  
From the view of spin systems the principal gain here is the
ability to describe the latter, as a direct analysis in the spin-1/2
case is plagued by Berry phases.  Nevertheless, even in the
case of integer-spin systems, the dual approach leads to interesting and 
unexpected predictions as mentioned above.

The formulation of a low-energy dual theory proceeds 
schematically as follows.  Let us focus on the easy-plane triangular
lattice antiferromagnet (either integer-spin or spin-1/2), which 
dualizes to a system of half-filled bosonic
vortices on the honeycomb lattice with ``electromagnetic'' interactions.
First, the bosonic vortices are converted into fermions carrying $2\pi$ 
fictitious flux via a Jordan-Wigner transformation\cite{JordanWigner}.  
At this stage, the dual theory
describes fermionic vortices hopping on the dual lattice, coupled to a
Chern-Simons field and $U(1)$ gauge field that mediates logarithmic
vortex interactions.  Next, one considers a mean-field state in which
fluctuations in the gauge fields are ignored and the flux is taken as 
an average background.  
Such ``flux-smeared'' treatments have been successful in the
composite-particle approach to the fractional quantum Hall 
effect.\cite{Heinonen}
Pursuing a similar approach in the context of spin systems, 
Lopez \emph{et al}.\cite{Lopez} developed a theory of the square-lattice 
Heisenberg antiferromagnet in terms of fermions coupled to a 
Chern-Simons field. 
Compared with the latter application, we have an advantage working
with vortices as there is an additional $U(1)$ gauge
field which gives rise to long-range interactions that strongly
suppress vortex density fluctuations.  This fact will play a 
central role in our analysis.  Exchange statistics are consequently 
expected to play a lesser role compared with the
case of weakly-interacting fermions.  Replacing the Chern-Simons 
flux by its average should then be a good starting point for 
deriving a low-energy theory, and the resulting continuum theory should
be more tractable.  

It is straightforward to diagonalize the ``flux-smeared'' 
mean-field hopping Hamiltonian, which at vortex half-filling reveals 
two Dirac nodes in the integer-spin case and four Dirac nodes in the 
spin-1/2 case.  
Focusing on excitations in the vicinity of these nodes and restoring 
the gauge fluctuations, one finally arrives at a low-energy Dirac theory 
with either two or four species of fermions coupled to the Chern-Simons 
and $U(1)$ gauge fields.  
In the absence of the Chern-Simons field, 
the theory becomes identical to QED3.

In the spin-1/2 case, this dual theory has a very rich structure, and
describes either a critical point in the spin model or a 
new critical spin-liquid phase.  As mentioned above, we defer an
analysis of spin-1/2 systems to a subsequent paper, focusing instead on
the integer-spin case.  Integer-spin systems provide an ideal setting
for us to establish the dual approach introduced here, as much is known about
their critical behavior from a direct analysis of the spin model.  
Viewed another
way, using the known results from a direct approach leads us to
the remarkable predictions described above for the dual critical
theory, which are highly nontrivial from the latter perspective.  
We proceed now by
providing a more detailed summary of the dual analysis for the 
integer-spin case, beginning with the spin model to make the preceding
discussion more concrete.

Throughout the paper, the easy-plane spins are cast in terms of
quantum rotors.  The integer-spin Hamiltonian with nearest-neighbor 
interactions then reads
\begin{equation}
  H_{XY} = J\sum_{\langle {\bf r r'}\rangle} 
            \cos(\varphi_{\bf r}-\varphi_{\bf r'}) 
  + U \sum_{\bf r} n_{\bf r}^2,
  \label{Hxyintro}
\end{equation}
where $n_{\bf r}$ is the density, $\varphi_{\bf r}$ is the conjugate
phase, and $J,U>0$.  For small $U/J$ [i.e., low temperatures in the
corresponding (2+1)D classical system], this model realizes 
the well-known $\sqrt{3}\times\sqrt{3}$ coplanar ordered phase,
although deformations of the model can give rise to collinear magnetic 
order as well.  For large $U/J$ the paramagnetic phase arises.  
The dual vortex Hamiltonian obtained from Eq.\
(\ref{Hxyintro}) can be conveniently written 
\begin{eqnarray} 
  {\mathcal H}_{\text{dual}} &=& -2 t_v \sum_{\langle {\bf x x'} \rangle}
  \cos(\theta_{\bf x}-\theta_{\bf x'} -a_{\bf x x'})
  \nonumber \\ 
  &+& \sum_{\bf x x'}(N_{\bf x}-1/2) V_{\bf x x'}(N_{\bf x'}-1/2)
  \nonumber \\
  &+& U' \sum_{\bf r}(\Delta \times a)_{\bf r}^2
     + J' \sum_{\langle {\bf x x'} \rangle} e_{\bf x x'}^2 , 
  \label{dualHintro} 
\end{eqnarray}
with $J',U'>0$.  Here $a_{\bf x x'}\in{\mathbb R}$ is a gauge field living on the 
links of the dual honeycomb lattice and $e_{\bf x x'}$ is the conjugate electric
field, $e^{i\theta_{\bf x}}$ is a vortex creation operator, and $N_{\bf x}$
is the vortex number operator.  $(\Delta \times a)_{\bf r}$
in the last line denotes a
lattice curl around each hexagon of the dual lattice.  The $t_v$ term
allows nearest-neighbor vortex hopping, and $V_{\bf x x'}$ encodes the 
logarithmic vortex interactions.  This form of the dual Hamiltonian
makes manifest the half-filling of the vortices.

To proceed, the vortices are converted into fermions coupled to
a Chern-Simons gauge field.
Because the vortices are half-filled, the Chern-Simons flux through
each dual plaquette averages to $2\pi$, which on the lattice is
equivalent to zero flux.  
Hence, in the flux-smeared mean-field state we simply set the 
gauge fields to zero.  We emphasize again that this is expected to
be a good approximation due to the vortex interactions, which
strongly suppress density fluctuations.  One then has
fermions at half-filling hopping on the honeycomb lattice.  
Diagonalizing the resulting hopping Hamiltonian, one finds two Dirac
nodes.  These nodes occur at the \emph{same wavevectors}
-- namely, the $\sqrt{3} \times \sqrt{3}$ ordering wavevectors --
that characterize the low-energy physics in the Landau
approach to the spin model, which are not apparent in the dual
theory of bosonic vortices but arise naturally in the fermionized
representation.  
A low-energy description can be obtained by focusing only on
excitations near these nodes, and upon restoring gauge fluctuations
one finds the following Euclidean Lagrangian density,
\begin{eqnarray} 
  {\mathcal L} &=& 
  \barpsi_a \gamma^\mu (\partial_\mu - i a_\mu - i A_\mu )\psi_a 
  \nonumber \\
  &+& \frac{1}{2 e^2} (\epsilon_{\mu\nu\lambda}\partial_\nu a_\lambda)^2 
  + \frac{i}{4\pi} \epsilon_{\mu\nu\lambda}A_\mu \partial_\nu A_\lambda 
  + {\mathcal L}_{\rm 4f}.
  \label{Lintro} 
\end{eqnarray} 
Here $\psi_{R/L}$ denote the two-component spinors corresponding to each
node, the flavor index $a = R/L$ is implicitly summed, 
and $A$ is the Chern-Simons field.  The usual form for the Chern-Simons
action above enforces the flux attachment that restores the bosonic
exchange statistics to the vortices.  Finally, 
${\mathcal L}_{\rm 4f}$ represents four-fermion terms arising from
short-range interactions allowed in the microscopic model.

The phases of the original spin model correspond to massive phases 
for the fermions in this formulation.  The paramagnetic phase 
is realized in the dual fermionic language as an integer 
quantum Hall state, with gapped excitations corresponding to the 
magnon of the paramagnet.  The magnetically-ordered states correspond 
to spontaneous mass generation driven by fermion interactions;
the $\sqrt{3}\times\sqrt{3}$ spin state is obtained as a vortex
``charge-ordered'' state, while the collinearly-ordered states 
are realized as vortex valence bond solids.  

Since the fermionized dual formulation faithfully captures the phases 
of the original spin model, it is natural to ask about the phase 
transitions in this framework.  Our exploration of criticality 
begins by implementing the following trick.  First, note that the fermions 
couple to the combination $\tilde a_\mu = a_\mu + A_\mu$.  
Rewriting the Lagrangian in terms of this sum field, we then integrate 
out the Chern-Simons field to arrive at a Lagrangian of the form
\begin{eqnarray} 
 {\mathcal L} &=& \barpsi_a
                  \gamma^\mu (\partial_\mu - i \tilde{a}_\mu )\psi_a
  + \frac{1}{2 e^2} (\epsilon_{\mu\nu\lambda}\partial_\nu
                  \tilde{a}_\lambda)^2
  \nonumber \\ 
  &+& {\cal L}_{4f} 
  + i \frac{\pi}{e^4} (\nabla \times \tilde{a}) \cdot  
          (\nabla \times \nabla \times \tilde{a}) ~.
\label{Lfinalintro}
\end{eqnarray} 
Remarkably, the lowest-order term generated by this procedure is the last
Chern-Simons-like term containing three derivatives, which 
at criticality is irrelevant by power counting compared to the 
Maxwell term.  It is therefore tempting to
drop this term altogether, which is tantamount to discarding the
bosonic exchange statistics and replacing the
bosonic vortices by fermions without any statistical flux
attachment, or more precisely a complete 
`screening' of the statistical gauge field.  
Such a replacement leads to an additional symmetry
corresponding to the na\"{\i}ve time reversal for fermions 
\emph{without} any Chern-Simons field, a symmetry which the
higher-derivative term violates.  The unimportance of exchange
statistics and concomitant emergence of this additional time-reversal 
symmetry at criticality seems plausible on physical grounds since the
logarithmic vortex interactions strongly suppress density
fluctuations.  Indeed, the trick performed above is only possible 
because of the additional $U(1)$ gauge field that mediates these 
interactions.  
We postulate that it is indeed legitimate at criticality to drop this
higher-derivative term, and that the resulting QED3 theory,
\begin{eqnarray} 
  {\mathcal L}_{QED3} = \barpsi_a 
                  \gamma^\mu (\partial_\mu - i \tilde{a}_\mu )\psi_a 
  + \frac{1}{2 e^2} (\epsilon_{\mu\nu\lambda}\partial_\nu
                  \tilde{a}_\lambda)^2
  + {\cal L}_{4f}, 
\label{QED3intro}
\end{eqnarray} 
contains a description of criticality in the spin model.  
Furthermore, we boldly conjecture that the critical QED3 theory
with the full $SU(2)$ symmetry describes the $O(4)$ multicritical 
point of the spin model where the paramagnet and two magnetically-ordered 
phases meet.

We support the above conjecture by establishing the
symmetry-equivalence of operators in the spin model with operators in QED3.
Specifically, we identify the two complex order parameter fields of the spin
model with the two leading monopole insertions of QED3, which add discrete units
of $2\pi$ gauge flux.  Moreover, we 
establish the equivalence of the various spin bilinears 
with fermionic bilinears and double-strength monopole
operators of QED3.  At the multicritical point, all of the spin
bilinears (except the energy) must have the same scaling dimension due to the emergent
global $O(4)$ symmetry.  This leads us to another bold prediction:
specific fermionic bilinears and double-strength monopole insertions in QED3
with two fermion species have the same scaling dimension.  
We again remark that this would be quite surprising, as from the
perspective of the dual theory there is
\emph{a priori} no apparent relation between these operators.

The above predictions are in principle verifiable in lattice QED3 
simulations.\cite{Kogut}  However, this likely requires fine-tuning of the 
lattice model to criticality, since we expect that the system 
generically ends up in a phase with spontaneously-generated mass 
driven by the fermion interactions that become relevant for the
low number of fermion flavors in our theory.  In particular, based on the 
correspondence with the $O(4)$ fixed point, we expect two 
four-fermion interactions to be
relevant at the QED3 fixed point (one strongly relevant and one weakly
relevant).  We note that we have performed a
leading-order ``large-$N$'' analysis to assess the stability of the QED3 fixed 
point in the limit of a large number of fermion flavors, where gauge
fluctuations are suppressed.  This analysis suggests that 
only one interaction in the two-flavor theory is relevant;
however, higher-order corrections could certainly dominate and lead to
two relevant interactions as predicted above.

\vskip 2mm
We now want to give a physical view of the fermionized vortices and 
also point out a connection with the so-called dipole interpretation 
of a Fermi-liquid-like state of bosons at
filling factor $\nu=1$.\cite{PH,Read}  This system was studied
in connection with the Halperin, Lee, and Read (HLR) theory\cite{HLR}
of a compressible fractional quantum Hall state for electrons at $\nu=1/2$.
In the direct Chern-Simons approach, the two systems lead
to very similar theories, but many researchers have argued
the need for improvement over the HLR treatment 
(for a recent review see Ref.~\onlinecite{MS}).
A detailed picture for bosons at $\nu=1$ is developed in 
Refs.~\onlinecite{Read,PH} working in the lowest Landau
level.  The present approach of fermionizing vortices rather than the
original bosons provides an alternative and perhaps more 
straightforward route for describing this system and obtaining 
the results of Ref.~\onlinecite{Read}.
Our approach is closest in spirit to Ref.~\onlinecite{DHL}, 
which also uses duality to represent vortices.
The fermionized vortices appear to be `natural' fermionic degrees of 
freedom for describing the compressible quantum Hall state.  We also
note an early work\cite{Feigelman} where the qualitative possibility of
fermionic vortices was raised in a context closer to frustrated
easy-plane magnets.

The vortex fermionization procedure is essentially the same as for
the spin model described above.  More details for this case can be found in 
Appendix~\ref{app:nu1}.  Dualizing the $\nu = 1$ bosons, one obtains a
theory of bosonic vortices at filling factor $\nu_{\rm dual} = 1/\nu =
1$, with mean vortex density equal to the density of the original
bosons $\bar\rho$.  
Again, unlike the original bosons, the vortices interact via a 2D 
electromagnetic interaction, and fermionization of vortices
gives a more controlled treatment.
We obtain fermions in zero average field, coupled to a $U(1)$ 
gauge field, $a$, and a Chern-Simons gauge field, $A$.
This theory is very similar to Eq.~(\ref{Lintro}), but with
nonrelativistic fermions at density $\bar\rho$.
With this theory, one can, for example, reproduce the physical response 
properties expected of the compressible state\cite{Read} 
by using an RPA approximation for integrating out the fermions.

It is useful to exhibit the equivalent of Eq.~(\ref{Lfinalintro}) 
in the present case,
\begin{eqnarray} 
{\mathcal L}_{\nu=1} &=& 
\psi^\dagger\left(\partial_\tau 
                  -\frac{[{\bm \nabla} - i \tilde{\bm a}]^2}{2 m_{\rm vort}}
       \right) \psi 
- i \tilde{a}_0 (\psi^\dagger\psi - \bar\rho) + \dots~.
\end{eqnarray}
Here $m_{\rm vort}$ is a phenomenological vortex mass.  As before,
$\tilde a = a + A$, and the Chern-Simons field has been integrated
out.  
The final action has nonrelativistic fermions coupled to the gauge field 
$\tilde{a}$.   As detailed in Appendix~\ref{app:nu1}, in the
lowest Landau level limit there is \emph{no} Chern-Simons like term
for the field $\tilde{a}$ in contrast to Eq.~(\ref{Lfinalintro}), and the theory has the same structure as in Ref.~\onlinecite{Read}.

In the present formulation, the fermions are neutral and do not
couple directly to external fields.  However, their dynamics
dictates that they carry \emph{electrical dipole moments} oriented 
perpendicular to their momentum ${\bm k}$ and of strength proportional 
to $|{\bm k}|$, which can be loosely viewed as a dumbbell formed by an original 
boson and a vortex.
This can be seen schematically as follows.  First we note that the flux of 
$\tilde{\bm a}$ is equal to the difference in the boson and vortex densities:
\begin{equation}
{\bm\nabla} \wedge \tilde{\bm a}
= {\bm\nabla} \wedge {\bm a} + {\bm\nabla} \wedge {\bm A}
= 2\pi (\rho_{\rm bos} - \rho_{\rm vort}) ~.
\end{equation}
Here ${\bm\nabla} \wedge \tilde{\bm a}=
\epsilon_{ij} \partial_i \tilde{a}_j$ denotes a 2d scalar curl; we also define
$(\wedge {\bm v})_i = \epsilon_{i j} v_j$ for a 2d vector ${\bm v}$.
Consider now a wavepacket moving with momentum ${\bm k}$ in the $\hat{\bf y}$ 
direction, and imagine a finite extent in the $\hat{\bf x}$ direction.  
In the region of the wavepacket, we have $\tilde{a}_y = |{\bm k}|$, while 
$\tilde{a}_y$ is zero to the left and to the right of the wavepacket.  
Then ${\bm\nabla} \wedge \tilde{\bm a} \sim \partial_x \tilde{a}_y$ is 
positive near the left edge and negative near the right edge, i.e. the
boson density is higher than the vortex density on the left side
of the moving fermion and vice versa on the right side.
A more formal calculation is presented in Appendix~\ref{app:nu1}
and gives the following \emph{dipolar} expression for the original boson 
density in terms of the fermionic fields:
\begin{equation}
\delta\rho_{\rm bos} ({\bm q}) = i {\bm q} \cdot 
\int \frac{d^2 k}{(2\pi)^2} \; \frac{\wedge {\bm k}}{2\pi\bar\rho} \; 
     \psi^\dagger_{{\bm k}-{\bm q}/2} \psi_{{\bm k}+{\bm q}/2} ~.
\end{equation}

Thus, by fermionizing the vortices we indeed obtain a description of the
compressible $\nu=1$ state very similar to the dipole picture,
and this is achieved in a rather simple way without using the lowest 
Landau level projection.  The success of our approach in this case
makes us more confident in the application to frustrated
spin systems.

The remainder of the paper focuses on the integer-spin triangular
antiferromagnet and is arranged as follows.  In Sec.\ \ref{sec:DirectBoson} we 
review the phases and critical behavior that arise from a
direct analysis of the spin model.  The duality mapping is introduced
in Sec.\ \ref{sec:BosVort}.  In Sec.\ \ref{sec:FermVort} the 
fermionized-vortex formulation is developed.  A low-energy
continuum theory is obtained here, and we also discuss how to
recover the phases of the spin model from the fermionic picture.  
An analysis of the QED3 theory conjectured to describe criticality in the
spin model is carried out in Sec.\ \ref{sec:QED}.  We conclude with a
summary and discussion in Sec.\ \ref{sec:Conclusion}.

\section{Direct Landau Theory}
\label{sec:DirectBoson}

\subsection{The Model}
\label{subsec:model}
 
In this paper we focus on quantum triangular XY antiferromagnets with 
integer spin, described by the Hamiltonian
\begin{equation}
  H = J/2\sum_{\langle {\bf r r'}\rangle} 
        (S_{\bf r}^+ S_{\bf r'}^- + S_{\bf r}^- S_{\bf r'}^+) 
    + U \sum_{\bf r} (S_{\bf r}^z)^2,
\end{equation}
with $J,U>0$, $S_{\bf r}^{\pm} = S_{\bf r}^x \pm i S_{\bf r}^y$,
and $[S_{\bf r}^\alpha , S_{\bf r'}^\beta] =  i \delta_{\bf r r'} 
\epsilon_{\alpha \beta \gamma} S_{\bf r}^\gamma$.
Rather than working with spin models, it will be particularly convenient
to instead model such easy-plane systems with rotor variables, 
introducing an integer-valued field $n_{\bf r}$ and a $2\pi$-periodic 
phase variable $\varphi_{\bf r}$ that satisfy 
$[\varphi_{\bf r}, n_{\bf r'}] = i \delta_{\bf r r'}$.  
Upon making the identification, $S_{\bf r}^+ \to e^{i \varphi_{\bf r}}$
and $S_{\bf r}^z \to n_{\bf r}$, the appropriate rotor XY Hamiltonian 
reads,
\begin{equation}
  H_{XY} = J\sum_{\langle {\bf r r'}\rangle} 
            \cos(\varphi_{\bf r}-\varphi_{\bf r'}) 
  + U \sum_{\bf r} n_{\bf r}^2.
  \label{Hxy}
\end{equation}
Equation (\ref{Hxy}) can equivalently be written 
\begin{equation}
  H_{XY} = -J \sum_{\langle {\bf r r'}\rangle} 
              \cos(\phi_{\bf r}-\phi_{\bf r'} + {\cal A}^0_{\bf r r'}) 
  + U \sum_{\bf r} n_{\bf r}^2,
  \label{Hxy2}
\end{equation}
where we have introduced new phase variables $\phi_{\bf r}$ and a 
static vector potential ${\cal A}^0_{\bf r r'}$ that induces 
$\pi$ flux per triangular plaquette.  
In this form, the nearest-neighbor
couplings can be viewed as ferromagnetic, with frustration arising
instead from the flux through each plaquette.  The model can 
alternatively be viewed as bosons at integer-filling hopping on the 
triangular lattice in an external field giving rise to one-half flux 
quantum per triangular plaquette.

A closely related model also described by Eq.~(\ref{Hxy2}) is the
fully-frustrated XY model on a square lattice, which has applications
to rectangular Josephson junction arrays.  The techniques applied to
the triangular antiferromagnets in this paper can be extended in a 
straightforward manner to this case as well.

Within a Euclidean path integral representation, the 2D quantum model 
becomes identical to a vertical stacking of classical triangular 
XY antiferromagnets with ferromagnetic XY coupling between successive 
planes, imaginary time giving one extra classical dimension.  
This classical model has been extensively studied, 
and much is known about the phases and intervening phase transitions.  
We will be particularly interested in the phase transitions,
which are quantum transitions in the spin model.
In this section we will briefly review the known results obtained from
a Landau-Ginzburg-Wilson analysis and from Monte Carlo studies.
This summary of the direct approach to these models will later serve 
as a useful counterpart to our dual approach.

\subsection{Low-energy Landau Theory}
\label{subsec:DirectLGW}
A low-energy effective theory for the triangular XY antiferromagnet can
be obtained by introducing a Hubbard-Stratonovich transformation that 
replaces $e^{i \varphi_i}$ by a ``soft spin'' $\Phi_i^*$ with the same
correlations.  This leads to a Euclidean action of the form
\begin{eqnarray}
  S &=& \int_\tau \bigg{[}\sum_{\bf r}(|\partial_\tau \Phi_{\bf r}|^2 
  + m |\Phi_{\bf r}|^2 + u|\Phi_{\bf r}|^4) \nonumber \\
  &+& {\mathcal J}\sum_{\langle {\bf r r'}\rangle}(\Phi_{\bf r}^*
  \Phi_{\bf r'} + c.c.)\bigg{]},
\end{eqnarray}
with ${\mathcal J}>0$.  [Note that we used the angle variables 
appearing in Eq.~(\ref{Hxy}) to implement this transformation.] 
Diagonalizing the kinetic part of the action in momentum space, 
one finds two global minima at wavevectors $\pm {\bf Q}$, 
where ${\bf Q} = (4\pi/3, 0)$.
The low-energy physics can be captured by expanding the fields around 
these wavevectors by writing
\begin{equation}
  \Phi ({\bf r}) \approx e^{i {\bf Q}\cdot {\bf r}}\Phi_R({\bf r}) 
                 + e^{-i {\bf Q} \cdot {\bf r}}\Phi_L({\bf r}),
\end{equation}
where $\Phi_{R/L}$ are complex fields assumed to be slowly-varying 
on the lattice scale.  

To explore the phases of the system within a Landau-Ginzburg-Wilson 
framework, one needs to construct an effective action for these fields 
that respects the microscopic symmetries of the underlying lattice model.
In particular, the action must preserve the following discrete
lattice symmetries:
translation by triangular lattice vectors $\delta {\bf r}$ 
($T_{\delta \bf r}$), $x$-reflection 
[${\cal R}_{x}: (x,y) \to (-x,y)$], 
and $\pi/3$ rotation about a lattice site ($R_{\pi/3}$).  
Additionally, the action must be invariant under a number of internal 
symmetries, specifically a global $U(1)$ symmetry,
$\varphi_{\bf r} \to \varphi_{\bf r} + \alpha$;
a ``particle-hole" or ``charge" conjugation (${\mathcal C}$) symmetry,
\begin{equation}
  {\mathcal C} \;:\quad  n_{\bf r} \rightarrow - n_{\bf r} , \quad 
  \varphi_{\bf r} \rightarrow - \varphi_{\bf r} ;
\end{equation}
and spin time reversal ($\mathcal T$) which sends 
$S_{\bf r}^\alpha \rightarrow -S_{\bf r}^\alpha$,
\begin{equation}
  {\mathcal T} \;:\quad  n_{\bf r} \rightarrow - n_{\bf r} , \quad
  \varphi_{\bf r} \rightarrow  \varphi_{\bf r} + \pi , \quad
  i \rightarrow -i .
\end{equation}
Under these symmetry operations, the continuum fields $\Phi_{R/L}$ 
transform according to
\begin{eqnarray}
  T_{\delta {\bf r}} &:& \Phi_{R/L} \rightarrow e^{\pm i {\bf Q}\cdot 
  {\delta {\bf r}}}\Phi_{R/L}, 
  \nonumber \\
  {\cal R}_x, R_{\pi/3} &:& \Phi_R \leftrightarrow \Phi_L, 
  \nonumber \\
  U(1) &:& \Phi_{R/L} \rightarrow e^{i \alpha} \Phi_{R/L}, 
  \\
  {\mathcal C} &:& \Phi_{R/L} \rightarrow \Phi_{L/R}^*, 
  \nonumber \\
  {\mathcal T} &:& \Phi_{R/L} \rightarrow \Phi_{R/L}^*. 
  \nonumber
\end{eqnarray}
In the last line above, we followed time reversal by a $U(1)$ transformation
to remove an overall minus sign.

These symmetry constraints lead to a continuum action of the form 
\begin{eqnarray} 
  S_{eff} &=& \int_{{\bf r},\tau} \bigg{ \{ }  
  \sum_{a = R/L} (|\partial_\mu \Phi_a|^2 + r |\Phi_a|^2) 
  \nonumber \\ 
  &+& u_4 (|\Phi_R|^2 + |\Phi_L|^2)^2 
   + v_4 |\Phi_R|^2 |\Phi_L|^2  
  \nonumber \\ 
  &-& v_6 [(\Phi_R^* \Phi_L)^3 + c.c.)]\bigg{ \} }. 
  \label{Seff} 
\end{eqnarray} 
The $v_6$ term has been included since it is the lowest-order term that 
reduces the symmetry down to the global $U(1)$ of the microscopic model.
Note that when $v_4 = v_6 = 0$ the action acquires an $O(4)$ symmetry. 
On the other hand, when $v_6 = 0$ and $v_4 = -2u_4$ the action reduces 
to two decoupled $U(1)$ models.  

We now proceed to explore the phases that arise from Eq.~(\ref{Seff}) 
within mean-field theory, assuming $u_4>0$ but allowing $r$, $v_4$, and 
$v_6$ to be either positive or negative.

\subsection{Phases}
\label{subsec:DirectPhases}
 
1. The simplest phase arises when $r>0$ so that 
$\langle \Phi_a \rangle = 0$.  
This corresponds to the quantum paramagnet with 
$\langle S^- \rangle = 0$ at each site. 
 
\begin{figure} 
  \begin{center} 
    {\resizebox{8cm}{!}{\includegraphics{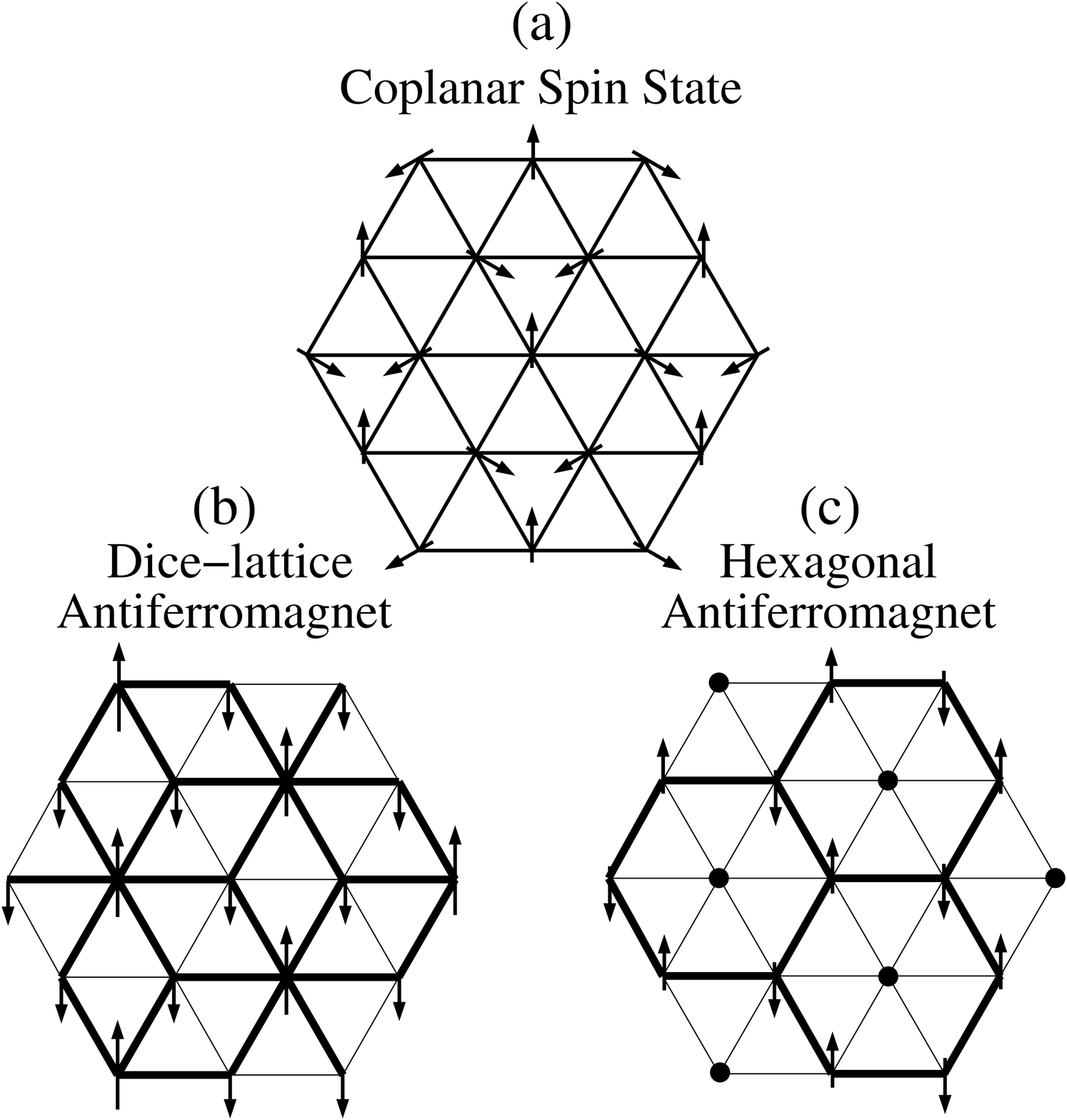}}} 
  \end{center} 
  \caption{(a) Coplanar spin configuration in the $\sqrt{3} \times 
    \sqrt{3}$ ordered state.  
    (b) Dice-lattice collinear antiferromagnetic state.  Bold lines 
    denote satisfied bonds, which form a dice lattice, while the
    remaining bonds are frustrated.  $|\langle S^- \rangle|$ is roughly 
    twice as large on sites with long arrows than on sites with short 
    arrows.  
    (c) Hexagonal collinear antiferromagnetic state.  Circles 
    represent fluctuating spins with $\langle S^- \rangle = 0$. 
    Satisfied bonds denoted by bold lines form a hexagonal lattice.} 
  \label{SpinSts} 
\end{figure}

When $r<0$, magnetic order develops, with the character depending 
on the sign of $v_4$: 
 
2. The case $r<0$, $v_4>0$ favors a state with  
$\langle \Phi_R \rangle \neq 0$ and $\langle \Phi_L \rangle = 0$  
(or vice versa).  Taking $\langle \Phi_R \rangle = 1$, the order  
parameter in terms of the original spin variables is $\langle 
S_{\bf r}^- \rangle = e^{i {\bf Q}\cdot {\bf r}}$.  This corresponds to  
the $\sqrt{3} \times \sqrt{3}$ state with $120^\circ$ coplanar order, 
whose spin configuration is illustrated in Fig.\ \ref{SpinSts}(a).   
 
3. When $r, v_4<0$, states with  
$\langle \Phi_{R/L}\rangle \sim e^{i\theta_{R/L}} \neq 0$ emerge.  
These states are characterized by collinear spin order,  
which can be obtained from 
\begin{equation} 
  \langle S_{\bf r}^- \rangle \sim e^{i\theta_+/2} 
  \cos({\bf Q}\cdot {\bf r} + \theta_-/2), 
\end{equation} 
where $\theta_\pm \equiv \theta_R \pm \theta_L$.  Two types of collinear 
spin states can occur depending on the phase difference $\theta_-$, 
which is determined by the sign of $v_6$.   
(i) For $v_6>0$, the action is minimized with a phase difference of 
$\theta_- = 2n\pi/3$, where $n$ is an integer.  The resulting spin order 
on the three sublattices of the triangular lattice can be written 
schematically as $(1, -\frac{1}{2}, -\frac{1}{2})$.   
This can be viewed as the dice-lattice collinear antiferromagnetic state 
depicted in Fig.~\ref{SpinSts}(b), where ``happy'' nearest-neighbor 
bonds with spins aligned antiparallel form a dice lattice as shown by 
the solid lines.  There are three distinct such ground states 
(in addition to broken global $U(1)$ symmetry) corresponding to the 
three inequivalent values of $\theta_-$. 
(ii) When $v_6<0$, a phase difference of $\theta_- = (2n+1)\pi/3$ is  
preferred.  The spin order on the three sublattices is then $(1, -1, 0)$, 
which can be viewed as hexagonal collinear antiferromagnetic order  
as illustrated in Fig.\ \ref{SpinSts}(c). 
In this state, spins on one of the three hexagonal subnetworks of  
the triangular lattice order antiferromagnetically, 
while spins on the remaining sites fluctuate around zero average.
Here also there are three distinct ground states arising from the 
inequivalent values of $\theta_-$. 
 
One can also distinguish between the coplanar and collinear 
magnetic orders by considering \emph{chirality} and 
\emph{bond energy wave} operators.
Spin chirality $\kappa_p$ at each triangular plaquette $p$ can be 
defined by  
\begin{equation} 
  \kappa_p \sim \sum_{\langle{\bf r r'}\rangle \in p}  
  \!\!\!\!\!\!\!\!\!\!\!\!\!\circlearrowleft \:\: 
  \sin(\varphi_{\bf r'}-\varphi_{\bf r}), 
\end{equation} 
where the sum is over the links of the plaquette oriented 
counterclockwise.  A local chirality operator is obtained by
summing $\kappa_p$ over local ``up'' triangles. 
A bond energy wave operator may be defined as 
\begin{equation} 
  B_{{\bf rr}'} \sim e^{i{\bf Q} \cdot ({\bf r} + {\bf r'})}  
  \cos(\varphi_{\bf r} - \varphi_{\bf r'})~. 
  \label{SpinBondOrder} 
\end{equation} 
In terms of the continuum fields $\Phi_{R/L}$, we identify 
\begin{eqnarray} 
  \kappa &\sim& |\Phi_R|^2 - |\Phi_L|^2 \equiv K_z ~, \nonumber \\  
  B &\sim& \Phi_R^* \Phi_L \equiv (K_x + i K_y)/2 ~.
  \label{ChirVbs2Kvec}
\end{eqnarray} 
Here we have introduced a three-component vector field, 
\begin{equation} 
  \bm{K}= \Phi^*_a \bm{\tau}_{ab} \Phi_b ,
  \label{Kvec} 
\end{equation} 
with $\bm\tau = (\tau^x, \tau^y, \tau^z)$ a vector of Pauli matrices. 
In the coplanar phase, the vector $\bm{K}$ points in the $\pm K_z$ 
direction, while in the collinear phase it lies in the $(K_x, K_y)$ 
plane.  The degeneracy in the plane is lifted by the $v_6$ term,
leaving three possible ordering directions $\theta_-$ for each collinear
phase as discussed above.
 
\begin{figure} 
  \begin{center} 
    {\resizebox{8cm}{!}{\includegraphics{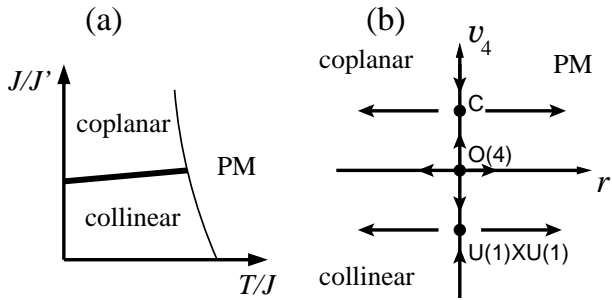}}} 
  \end{center} 
  \caption{(a) Sketch of the phase diagram in the vicinity of the 
    coplanar-collinear transition.  (b) Putative renormalization group 
    flows in the $(r,v_4)$ plane assuming a continuous transition 
    from both the coplanar and collinear states to the paramagnet.  } 
  \label{phases} 
\end{figure}

Microscopically, the coplanar, collinear, and paramagnetic phases 
can be realized in the classical stacked triangular XY antiferromagnet
with an additional ``boson pair hopping'' term, 
\begin{eqnarray} 
  H_{cl} &=&  -J\sum_{{\bf r},z} 
  \cos [\varphi_{{\bf r},z+1} - \varphi_{{\bf r},z}] \nonumber \\
  &+& J \sum_{\langle {\bf r r'} \rangle, z}  
  \cos[\varphi_{{\bf r},z}-\varphi_{{\bf r'},z}]\nonumber\\
   &-& J' \sum_{\langle {\bf r r'} \rangle, z}  
  \cos[2(\varphi_{{\bf r},z}-\varphi_{{\bf r'},z})], 
  \label{Hcl}
\end{eqnarray} 
where $J'>0$ and the $z$ coordinate labels the different triangular 
lattice planes. 
The $\sqrt{3} \times \sqrt{3}$ coplanar phase is the  
ground state when $J' \ll J$.  As $J'$ increases, collinear 
antiferromagnet order eventually becomes more energetically favorable.  
Of course, at high enough temperature we obtain the paramagnetic phase.
A sketch of the phase diagram containing the three phases is shown in 
Fig.\ \ref{phases}(a). 
 
\subsection{Critical Behavior}
\label{subsec:DirectCrit}
 
Significant numerical and analytical effort has been directed  
towards understanding the phase transitions occurring in classical 
stacked triangular XY antiferromagnets.  For a recent review, the 
reader is referred to Refs.\ \onlinecite{KawamuraRev, Vicari, NonpertRG, Calabrese}.
Here, we shall only highlight the known results that are pertinent to the 
present work.  A brief discussion of our own exploratory Monte Carlo
study of the 
classical model Eq.~(\ref{Hcl}) that realizes both the coplanar and 
collinear (hexagonal) spin-ordered phases will also be given.
 
In mean field theory, the paramagnet to magnetically-ordered 
transitions are continuous, while the transition from the coplanar to 
the collinear state is first order.   
 
The transition between the paramagnet and the collinear state
also appears to be continuous in our Monte Carlo study of the 
model Eq.~(\ref{Hcl}) near the multicritical point where the three phases meet.
Moreover, such a continuous transition is expected to be governed by 
a $U(1) \times U(1)$ fixed point with $v_4^* = -2u_4^*$ and  
$v_6^* = 0$.  Small deviations from the condition $v_4 = -2u_4$ 
give rise to energy-energy coupling of the two $U(1)$ models, which is 
irrelevant at the decoupled fixed point in three
dimensions.\cite{Vicari,Sak}
It should be noted that a $4-\epsilon$ expansion in the $v_4<0$ case  
predicts a continuous transition characterized by a \emph{different} 
stable fixed point, while the $U(1) \times U(1)$ fixed point is found 
to be unstable in that approach.\cite{KawamuraRG,Vicari}  Thus in this case
the $4-\epsilon$ expansion does not capture the critical physics in 
three dimensions.

The nature of the transition between the paramagnetic and coplanar 
phases is controversial.\cite{Vicari, NonpertRG, Calabrese}
Starting with Kawamura, several Monte Carlo simulations on the stacked 
triangular XY antiferromagnet conclude that the transition is 
continuous,\cite{KawamuraNum, tricritical, PhaseDiagram, biquadratic}
which also appears to be the case in our Monte Carlo simulations 
in the vicinity of the multicritical point.
However, recent simulations on larger lattices claim a very weak 
first-order transition,\cite{Itakura,Peles} and simulations of 
modified XY systems expected to be in the same universality class 
have also seen a first-order transition.\cite{modifiedXY,Peles}
On the other hand, Ref.~\onlinecite{Calabrese} claims a continuous
transition in the vicinity of the $O(4)$ fixed point in a model which
is a lattice discretization of the action Eq.~(\ref{Seff}).

Analytical results for the paramagnet-to-coplanar transition are 
conflicting as well.  For instance, recent 
six-loop perturbative renormalization group calculations in  
$d = 3$ dimensions\cite{6loop1, 6loop2} find a stable 
``chiral'' fixed point as conjectured by Kawamura\cite{KawamuraRG}.  
On the other hand, ``non-perturbative'' renormalization group 
techniques predict a weak first-order transition.\cite{NonpertRG}  
The $4-\epsilon$ expansion in the XY case 
($n=2$ case in Ref.\ \onlinecite{KawamuraRG}) finds runaway flows 
for $v_4>0$, which would be interpreted as predicting a first-order 
transition.

Finally, the multicritical point where the three phases meet
is described by an $O(4)$ fixed point with $v_4^*=v_6^*=0$; this
fixed point is unstable towards introducing a small nonzero $v_4$ term.
 
Figure \ref{phases}(b) contains a sketch of the renormalization group 
flows in the ($r, v_4$) plane expected if the transition from the 
paramagnet to the coplanar state is indeed continuous and governed by 
the corresponding fixed point, labeled ``C''.
(In the figure, the $u_4$ axis is perpendicular to the plane, and we are
considering fixed points that are stable in this direction.)
Fixed point C would disappear if this transition is first-order, 
leading to unstable flows toward $v_4 \to \infty$.
We note the possibility of observing crossover behavior controlled by 
the unstable $O(4)$ fixed point shown in Fig.\ \ref{phases}(b) by 
fine-tuning both $r$ and $v_4$. 
This can be explored with Monte Carlo simulations of $H_{cl}$ by 
observing scaling in the vicinity of the multicritical point  
where the paramagnet, coplanar, and collinear phases meet in 
Fig.\ \ref{phases}(a). 

\begin{table}[t]
\caption{\label{tab:phi} Transformation properties of the
bosonic fields and bilinears under the microscopic symmetries.  
Here $K_\pm = K_x \pm i K_y$ and $I_\pm = I_x \pm i I_y$.
For brevity, the lattice coordinates which should transform
appropriately under the lattice symmetries have been suppressed.
We also show field transformations under the modified reflection
$\tilde{\cal R}_x \equiv {\cal R}_x {\cal C} {\cal T}$, which
will be useful later.}
\begin{ruledtabular}
\begin{tabular}{c | c | c | c | c | c} 
  & $T_{\bf \delta r}$ & $R_{\pi/3}, {\mathcal R}_x $ & ${\mathcal C}$ 
  & ${\mathcal T}$ & $\tilde {\mathcal R}_x$ \\ 
  \hline
  $\Phi_{R/L} \to$  
  & $e^{\pm i {\bf Q}\cdot \delta{\bf r}} \Phi_{R/L}$
  & $\Phi_{L/R}$ 
  & $\Phi_{L/R}^*$
  & $\Phi_{R/L}^*$ 
  & $\Phi_{R/L}$ \\ 
  \hline
  $K_z \to$  
  & $K_z$ 
  & $-K_z$ 
  & $-K_z$  
  & $K_z$ 
  & $K_z$ \\ 
  \hline 
  $K_\pm \to$  
  & $e^{\mp 2i{\bf Q\cdot \delta r}} K_\pm$ 
  & $K_\mp$ 
  & $K_\pm$  
  & $K_\mp$ 
  & $K_\pm$ \\
  \hline 
  $I_z \to$  
  & $I_z$ 
  & $I_z$ 
  & $-I^*_z$  
  & $I^*_z$
  & $-I_z$ \\
  \hline 
  $I_\pm \to$  
  & $e^{\pm 2i{\bf Q\cdot \delta r}} I_\pm$ 
  & $-I_\mp$ 
  & $I^*_\mp$  
  & $I^*_\pm$ 
  & $-I_\pm$
  \end{tabular} 
\end{ruledtabular} 
\end{table} 
 
To examine the $O(4)$ symmetry at the multicritical point, it is 
convenient to express $\Phi_{R/L}$ as  
\begin{equation} 
  \Phi_R = \chi_1 + i \chi_2  ,  \quad
  \Phi_L = \chi_3 + i \chi_4  ,
  \label{O4vector} 
\end{equation} 
where $\chi_j$ are real fields.  
One expects that $\vec\chi \equiv (\chi_1, \chi_2, \chi_3, \chi_4)$ 
should transform as an $O(4)$ vector, and the ten independent 
bilinears $\chi_i \chi_j$ can be decomposed into an $O(4)$ scalar,
\begin{equation}
  K_0 \equiv \Phi_a^* \Phi_a = \vec\chi^2 ~,
  \label{K0}
\end{equation}
and a traceless, symmetric $4\times 4$ matrix that transforms as a 
second rank tensor representation of $O(4)$.  It will be useful later 
to observe that these nine bilinears can be arranged into the 
$SU(2)$ vector $\bm{K}$ defined in Eq.~(\ref{Kvec}), together with
two other $SU(2)$ vectors: 
\begin{equation} 
  \bm{I} = \Phi_a [\bm{\tau} \tau^y]_{ab} \Phi_b ,
\label{Ivec}
\end{equation} 
and its Hermitian conjugate, $\bm{I}^*$.  Under the global 
$U(1)$ symmetry one has $\bm{K} \to \bm{K}$
and $\bm{I} \to e^{2i\alpha} \bm{I}$.  For future comparisons
the transformation properties of these nine bilinears under the
remaining microscopic symmetries are shown in Table~\ref{tab:phi}.  
The $O(4)$ scalar $K_0$ defined in Eq.~(\ref{K0}) is invariant 
under all of the above microscopic symmetries.

One can furnish explicit microscopic realizations of the bilinears
$\bm{I}$ and $\bm{I}^*$, just as we did in Eq.~(\ref{ChirVbs2Kvec})
for the bilinears $\bm{K}$.  In our Monte Carlo study of the model
Eq.~(\ref{Hcl}), we monitored the apparent scaling dimensions 
of the nine bilinears and find that these indeed merge upon approaching 
the multicritical point, consistent with the expectation for the
$O(4)$ fixed point.

 
\section{Dual Vortex Theory}
\label{sec:BosVort}

\subsection{Duality Mapping}
\label{subsec:Dual}
 
\begin{figure} 
  \begin{center} 
    {\resizebox{5.5cm}{!}{\includegraphics{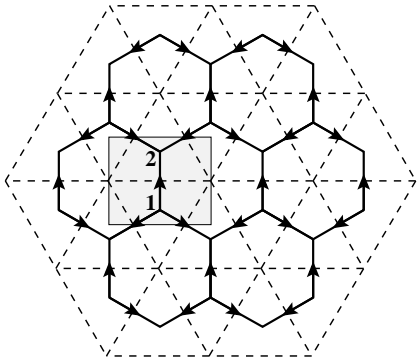}}} 
  \end{center} 
  \caption{Triangular lattice (dashed lines) and its 
    dual honeycomb lattice (solid lines).  The shaded region indicates 
    the two-site unit cell taken for the honeycomb, with the bottom 
    site labeled ``1'' and the top site ``2''. } 
  \label{dual} 
\end{figure}

In this section we introduce the dual formalism that will be used 
throughout the remainder of the paper.  At this point, it will prove 
convenient to work with the Hamiltonian as written in Eq.~(\ref{Hxy2}),
where frustration arises from the flux induced by the vector 
potential ${\mathcal A^0_{\bf r r'}}$.  We will implement the 
standard XY duality,\cite{duality} expressing the Hamiltonian 
in terms of gauge fields living on the links of the dual lattice, 
which in the case of the triangular lattice is the honeycomb 
(see Fig.\ \ref{dual}).  
 
As a first step, we define oriented gauge fields  
$\tilde e_{\bf x x'} \in [-1/2, 1/2)$ and  
$a_{\bf x x'} \in 2 \pi {\mathbb Z}$ living on the links of the 
honeycomb lattice, where ${\bf x}$ and ${\bf x'}$ label 
nearest-neighbor dual lattice sites.  
(We reserve ``${\bf x}$'' and ``${\bf r}$'' for sites of the honeycomb
and triangular lattices, respectively.)  The dual ``electric field'' 
$\tilde e_{\bf x x'}$ and ``vector potential'' $a_{\bf x x'}$ satisfy 
the commutation relation $[\tilde e_{\bf x x'}, a_{\bf x x'}] = i$ 
on the same link and commute on different links. 
 
The boson number and phase are related to the dual fields as follows: 
\begin{eqnarray} 
  n_{\bf r} &=& \frac{1}{2\pi}(\Delta \times a)_{\bf r}  
  \equiv \frac{1}{2\pi} \sum_{\langle{\bf x x'}\rangle \in p_{\bf r}}  
  \!\!\!\!\!\!\!\!\!\!\!\!\!\!\circlearrowleft \:\: a_{\bf x x'} 
  \label{a} \\ 
  e^{i \phi_{\bf r}} &=& \prod_{\bf r}^\infty  
  \!\!\!\!\!\!\!\!\!\!\longrightarrow  e^{2 \pi i \tilde{e}_{\bf x x'}}. 
  \label{e} 
\end{eqnarray}   
On the right side of Eq.\ (\ref{a}), the sum is over the 
counterclockwise-oriented links of the dual plaquette  
$p_{\bf r}$ enclosing site ${\bf r}$ of the triangular 
lattice.  The product in Eq.\ (\ref{e}) is carried 
out over an arbitrary string running along the triangular lattice  
from site ${\bf r}$ to spatial infinity.  A factor of  
$e^{2 \pi i \tilde{e}_{\bf x x'}}$ appears in the product for every 
link on the dual lattice bisected by the string, with ${\bf x}$ on the 
``right'' side of the string and ${\bf x'}$ on the ``left''.   
 
To ensure the uniqueness of the phases $\phi_{\bf r}$ obtained from 
Eq.\ (\ref{e}), the dual Hilbert space is constrained so that the 
operator 
\begin{equation} 
  N_{\bf x} \equiv (\Delta \cdot \tilde{e})_{\bf x} = 
  \sum_{\bf x'} \tilde{e}_{\bf x x'} 
  \label{N} 
\end{equation}  
has integer eigenvalues for all ${\bf x}$.  On the 
right-side of Eq.\ (\ref{N}), the sum is over the three nearest-neighbor 
sites of ${\bf x}$.  The meaning of ${N}_{\bf x}$ can be understood by 
using Eq.\ (\ref{e}) to show that 
\begin{equation} 
N_{\bf x} = \frac{1}{2\pi}  
  \sum_{\langle{\bf r r'}\rangle \in p_{\bf x}}  
  \!\!\!\!\!\!\!\!\!\!\!\!\!\!\circlearrowleft \:\: (\phi_{\bf 
  r'}-\phi_{\bf r})_{[-\pi,\pi)}, 
\end{equation} 
where the sum is over the counterclockwise-oriented links of the 
triangular plaquette $p_{\bf x}$ enclosing site ${\bf x}$ and 
the phase difference between adjacent sites is taken to lie in the 
interval $[-\pi,\pi)$.  The eigenvalues of this operator 
therefore encode \emph{vorticity}, i.e., the winding number of 
the phase around the triangular plaquette enclosing ${\bf x}$.   

To complete the duality transformation, define a static electric field  
$e^0_{\bf x x'}$ related to the vector potential  
${\cal A}^0_{\bf r r'}$ of Eq.~(\ref{Hxy2}) by 
\begin{equation} 
  e^0_{\bf x x'} = -\frac{1}{2\pi} {\cal A}^0_{\bf r r'}, 
  \label{e0} 
\end{equation} 
where $e^0_{\bf x x'}$ and ${\cal A}^0_{\bf r r'}$ live on 
intersecting links of the honeycomb and triangular lattices.  
Since ${\cal A}^0_{\bf r r'}$ generates $\pi$ flux per plaquette 
of the triangular lattice, we require $(\Delta \cdot e^0)_{\bf x}$ 
to be half-integer valued.  A convenient gauge choice we will now
employ is 
${\cal A}^0_{\bf r r'} = \pm \pi/3$ on each nearest-neighbor link, 
directed clockwise around ``up" triangles and counterclockwise 
around ``down" triangles.  In this gauge, $e^0_{\bf x x'} = 1/6$ along
the arrows shown in Fig.\ \ref{dual} so that 
$(\Delta \cdot e^0)_{\bf x} = +1/2$ on 
sublattice 1 while $(\Delta \cdot e^0)_{\bf x} = -1/2$ on sublattice 2. 
  
Upon defining $e_{\bf x x'} \equiv \tilde{e}_{\bf x x'} - e_{\bf x x'}^0$
we thereby arrive at a simple form for the dual Hamiltonian,  
\begin{equation} 
  {\mathcal H} = -J\sum_{\langle {\bf x x'} \rangle} \cos(2\pi 
  e_{\bf x x'}) + U' \sum_{\bf r}(\Delta \times a)_{\bf 
  r}^2, 
  \label{dualH} 
\end{equation} 
with $U' = U/(2\pi)^2$.  This Hamiltonian is supplemented by the 
constraint that,
\begin{equation}
 (\Delta \cdot e)_{\bf x} = N_{\bf x} - 1/2  ,
 \label{constraint}
\end{equation}
where we have found it convenient to shift the integer field 
$N_{\bf x} \rightarrow N_{\bf x} - 1$ for all ${\bf x}$ on 
sublattice 2.

As it stands, this Hamiltonian is difficult to work with due to the
integer constraint on the field $a_{\bf x x'}/(2\pi)$.  
It is therefore highly desirable to ``soften" this constraint,
allowing $a_{\bf x x'}$ to roam over all real numbers.  To be consistent 
$e_{\bf x x'}$ must also be taken on the real numbers, and it is then 
legitimate to expand the cosine term in Eq.\ (\ref{dualH}) to obtain,
\begin{eqnarray} 
  \tilde {\mathcal H} &=& \sum_{\langle {\bf x x'} \rangle}  
  [J' e_{\bf x x'}^2 -2 t_v  
  \cos(\theta_{\bf x}-\theta_{\bf x'} -a_{\bf x x'})]  
  \nonumber \\ 
  &+& U' \sum_{\bf r}(\Delta \times a)_{\bf r}^2, 
  \label{dualH2} 
\end{eqnarray}
with $J' \approx 2 \pi^2 J$.  
Here we have added a cosine term acting on the field $a_{\bf x x'}$ to 
implement the integer constraint softly, and explicitly expressed the
longitudinal piece of $a_{\bf x x'}$ as a lattice derivative of a 
``phase" field $\theta_{\bf x} \in [-\pi,\pi)$ living on the dual 
lattice sites.  The dual Hilbert space satisfies the constraint 
Eq.\ (\ref{constraint}) with the vortex number operator 
$N_{\bf x}\in {\mathbb Z}$ conjugate to $\theta_{\bf x}$. 
The operator $e^{i\theta_{\bf x}}$ creates a vortex at site ${\bf x}$ 
and satisfies the commutation relations 
$[N_{\bf x}, e^{i\theta_{\bf x'}}] 
 = \delta_{\bf x x'} e^{i\theta_{\bf x}}$.  
The $t_v$ term above thus allows nearest-neighbor vortex hopping in
addition to energetically favoring integer values 
for the density $n_{\bf r}$.  

Generically, we should also allow short-range vortex interaction
of the form
\begin{equation}
\sum_{\bf x x'} (N_{\bf x}-1/2) V^{\rm sr}_{\bf x x'} (N_{\bf x'}-1/2),
\end{equation}
in addition to the retarded long-range interaction mediated by the
gauge fields.  For a static configuration of vortices or in the
instantaneous limit when the ``speed of light'' $(U'/J')^{1/2}$ is 
infinite, the latter interaction becomes a 2D lattice Coulomb 
potential varying as 
$V^{\rm Coul}_{\bf x x'} \sim -J' \ln|{\bf x} - {\bf x'}|$.

\subsection{Phases in the Dual Vortex Variables}
\label{subsec:BosonicPhases}
Physically, the dual Hamiltonian describes vortices hopping on the sites
of the dual honeycomb lattice, interacting via a 2D ``electromagnetic" 
interaction.  Most importantly, the vortices are at
\emph{half-filling}, which can be directly traced to the frustration
on the original triangular lattice plaquettes.  This highlights
the distinction between the triangular lattice XY antiferromagnet 
and its unfrustrated counterparts, which have a dual vortex theory 
with an integer mean vortex number.

To illuminate the challenge in treating the strongly interacting vortex
system when at half-filling, it is instructive to consider how 
the various phases of the original model are described in terms 
of the dual variables.  The magnetically-ordered states 
correspond to ``insulating" phases of the dual vortices.  For 
example, the ordered phase with $120^\circ$ coplanar order of the 
XY spins corresponds to a ``vortex density wave" state in which the 
vortices sit preferentially on one of the two sublattices
of the dual honeycomb lattice.  On the other hand, the ordered states 
with collinear order correspond to ``vortex valence bond" phases;
one signature of these phases is the bond energy density order, which
can be measured both in terms of the vortices and in terms of the 
original spins.  Specifically, the bond energy wave operator
for the vortices takes a similar form as in Eq.~(\ref{SpinBondOrder}), 
except with the vortex bond energy operator inserted,
\begin{equation} 
  B_{{\bf rr}'} \sim e^{i{\bf Q} \cdot ({\bf r} + {\bf r'})}  
  \cos(\theta_{\bf x} - \theta_{\bf x'} - a_{\bf x x'}) ~.
  \label{VortexBondOrder} 
\end{equation} 
Here, the nearest-neighbor pairs ${\bf r}$, ${\bf r'}$ and ${\bf x}$,
${\bf x'}$ are bridged by intersecting links of the triangular and
honeycomb lattices.  

In all of these magnetically-ordered phases, since the dual vortices 
are ``insulating" and immobile, the ``photon" in the dual fields 
$a_{\bf x x'}$ and $e_{\bf x x'}$ can freely propagate.  This 
``photon" mode corresponds to the Goldstone spin-wave mode of 
the original XY spins.  On the other hand, the paramagnetic 
phase for the XY spins corresponds to a phase within which the 
dual vortices have condensed, 
$\langle e^{i \theta_{\bf x}} \rangle \ne 0$.  
In this dual vortex ``superfluid" the dual gauge flux is expelled, 
and the dual gauge field $a_{\bf x x'}$ picks up a Higgs mass;
thus, the spectrum is gapped in this phase as expected in the
spin paramagnet.

The quantum phase transitions separating the paramagnet from the coplanar
and collinear ordered spin states correspond to ``superfluid-insulator" 
transitions for the dual vortices.  But all of these vortex 
``insulators" involve a spontaneous breaking of lattice 
translational symmetries, due to the fact that the vortices are 
at half-filling.  Due to these complications it is not at all clear 
how one can possibly access the critical properties of these 
transitions from this dual vortex formulation.  In the next 
section we demonstrate how this can be achieved, by fermionizing 
the vortices.  Before undertaking this rather tricky business, we 
first summarize how the vortex fields transform under all of the 
lattice and internal symmetries.

\subsection{Symmetry Transformations for Vortices}
\label{subsec:BvortSym}
The transformation properties of the dual vortex and gauge fields can 
be deduced upon inspection of their defining equations, 
Eqs.~(\ref{a}) and (\ref{e}), together with the transformation properties
of the rotor fields $\phi_{\bf r}$ and $n_{\bf r}$.  
For example, lattice translations $T_{\bf \delta r}$ and $\pi/3$
rotations $R_{\pi/3}$ transform the
dual coordinates (e.g., ${\bf x}\rightarrow {\bf x} + {\bf \delta r}$
under $T_{\bf \delta r}$), but do not transform the dual gauge fields
$a, e$ or vortex fields $\theta, N$ as shown in Table \ref{tab:dual}.
Here and in what follows, our coordinate origin is always at a 
triangular lattice site.
Under charge conjugation ${\mathcal C}$, however, 
the fields $a, e, \theta$ change sign while $N\rightarrow 1-N$.  
Similarly, time reversal ${\mathcal T}$ changes the sign of the 
dual fields $a, \theta$ but leaves $e, N$ unchanged.  
Since ${\cal T}$ is accompanied by complex conjugation
the vortex creation operator $e^{i\theta}$ remains invariant under
time reversal.  

For lattice reflections it is convenient for later developments 
to henceforth consider a modified (antiunitary) transformation,
$\tilde{\cal R}_x$, which is defined as ${\cal R}_x$ combined with
${\cal C}$ and ${\cal T}$:
\begin{equation}
\tilde{\cal R}_x \equiv {\cal R}_x {\cal C} {\cal T} .
\end{equation}
The dual fields transform under $\tilde {\mathcal R}_x$ in the same
way they transform under time reversal, with the dual coordinates
additionally $x$-reflected.  
The complete set of symmetry transformations for the dual fields 
is summarized in Table \ref{tab:dual}.

Finally, we remark that the global $U(1)$ or XY spin symmetry corresponding 
to $e^{i\phi} \rightarrow e^{i(\phi+\alpha)}$ is not directly manifest 
in the dual vortex formulation, as can be seen by examining Eq.~(\ref{e}).  
This global $U(1)$ symmetry reflects the underlying conservation of
the boson number $n_{\bf r}$ (or the $S^z$-component of spin) 
and is replaced in the dual formulation by the conservation of the dual
flux $(\Delta \times a)_{\bf r}$.  
Note also that the dual theory has a $U(1)$ gauge redundancy 
(not a physical symmetry) given by 
$a_{\bf x x'} \to a_{\bf x x'} + \Lambda_{\bf x}-\Lambda_{\bf x'}$ 
and $\theta_{\bf x} \to \theta_{\bf x}+\Lambda_{\bf x}$, 
where $\Lambda_{\bf x}\in {\mathbb R}$.

\begin{table} 
\caption{\label{tab:dual} Transformation properties of the dual fields 
under the microscopic symmetries.  For brevity, the lattice coordinates, 
which also transform appropriately under the lattice symmetries, 
have been suppressed on all fields.  Here we use a modified reflection 
$\tilde {\mathcal R}_x$, which is a combination of ${\mathcal R}_x$,
${\mathcal C}$, and ${\mathcal T}$.  
Both ${\cal T}$ and $\tilde{\cal R}_x$ are accompanied with a complex 
conjugation, so that e.g. $e^{i \theta}$ is invariant.   } 
\begin{ruledtabular} 
\begin{tabular}{c | c | c } 
  $T_{{\bf \delta r}}$, $R_{\pi/3}$  
  &  ${\mathcal C}$ 
  & $\tilde {\mathcal R}_x$, ${\mathcal T}$ \\ 
  \hline 
  $a\rightarrow a$  
  & $a\rightarrow-a$ 
  & $a\rightarrow -a$ \\ 
  \hline 
  $ e \rightarrow e$ 
  & $e \rightarrow -e $ 
  & $ e\rightarrow e$ \\ 
  \hline 
  $\theta\rightarrow\theta$  
  & $\theta\rightarrow-\theta$ 
  & $\theta\rightarrow-\theta$ \\ 
  \hline 
  $N\rightarrow N$  
  & $N\rightarrow 1-N$  
  & $N\rightarrow N$
\end{tabular} 
\end{ruledtabular} 
\end{table}

\section{Fermionization of vortices}
\label{sec:FermVort}
 
\subsection{Chern-Simons Flux Attachment}
\label{subsec:CS}
As described in Sec.~\ref{subsec:BosonicPhases}, an analysis of the
phase transitions in the dual bosonic vortex theory is highly
nontrivial because the vortices are at half-filling, and we do not
know how to formulate a continuum theory of such bosonic degrees of
freedom.  We will sidestep difficulties associated with the finite
vortex density by \emph{fermionizing} the vortices, which as we will
demonstrate enables one to make significant progress in this direction.  
Specifically, we first treat the vortices as hard-core bosons, with the 
identification 
\begin{equation} 
  b_\bfx^\dagger \sim e^{i \theta_\bfx}, \quad  
  N_\bfx = b_\bfx^\dagger b_\bfx = 0, 1 ~. 
\end{equation} 
This is a reasonable approximation due to the repulsive vortex interactions, 
and the hard-core condition does not affect generic behavior.   
We then perform a two-dimensional Jordan-Wigner 
transformation\cite{JordanWigner}, 
\begin{eqnarray} 
  b_\bfx^\dagger &=& d_\bfx^\dagger 
  \exp[i \sum_{\bfx' \neq \bfx} \arg(\bfx, \bfx') N_{\bfx'} ], 
  \label{JW} 
  \\ 
  N_\bfx &=& b_\bfx^\dagger b_\bfx = d_\bfx^\dagger d_\bfx. 
\end{eqnarray} 
Here, $\arg(\bfx, \bfx')$ is an angle formed by the vector 
$\bfx - \bfx'$ with some fixed axis.  It is simple to check 
that $d_\bfx$ are fermionic operators satisfying the anticommutation  
relations $\{d_\bfx, d_{\bfx'}^\dagger\} = \delta_{\bfx \bfx'}$. 
The vortex hopping part of the Hamiltonian becomes 
\begin{eqnarray} 
-t_v \sum_{\la \bfx_1 \bfx_2 \ra}  
[ d_{\bfx_1}^\dagger d_{\bfx_2}  
  e^{-i A_{\bfx_1 \bfx_2}} e^{-i a_{\bfx_1 \bfx_2}} + \Hc ] ~.
  \label{VortFermHop} 
\end{eqnarray} 
The Chern-Simons field 
\begin{equation} 
A_{\bfx_1 \bfx_2} \equiv \sum_{\bfx' \neq \bfx_1,\bfx_2}  
[\arg(\bfx_2, \bfx') - \arg(\bfx_1, \bfx')] N_{\bfx'} 
\label{Adef}
\end{equation} 
lives on the links of the honeycomb lattice and is completely 
determined by the positions of the particles. 
In this transformation, we have essentially expressed hard-core bosons  
on the lattice as fermions each carrying a fictitious $2\pi$ flux.   
We used the Hamiltonian language in order to facilitate our discussion 
of the discrete symmetries below. 
 
The Chern-Simons field satisfies 
\begin{eqnarray} 
  (\Delta \times A)_{\bf r} =  
  \frac{2\pi}{3} \sum_{\bfx \in p_{\bf r}} (N_\bfx - 1/2),
\label{Aconstraint} 
\end{eqnarray} 
i.e., the fictitious $2\pi$ flux attached to a fermion can be viewed 
as equally-shared among its three adjacent plaquettes.   
In the above equation, we subtracted $1/2$ from each $N_\bfx$ 
to remove an unimportant $2\pi$ flux from each hexagon.  
This constitutes a convenient choice such that the Chern-Simons 
flux piercing the  
dual lattice vanishes on average since the vortices are half-filled.
There are further restrictions on the Chern-Simons field which we write  
schematically as $\Delta \cdot A = 0$ appropriate in the  
continuum limit. 
 
The complete Hamiltonian describes half-filled fermions with 
nearest-neighbor hopping on the honeycomb lattice, coupled to the 
$U(1)$ gauge field $a_{\bfx \bfx'}$ and the Chern-Simons 
field $A_{\bfx \bfx'}$.  
We can crudely say that the gauge field $a_{\bfx \bfx'}$ gives rise to 
repulsive logarithmic interactions between the fermions.

Under the microscopic symmetries, the fields in this fermionized 
representation transform as shown in Table~\ref{tab:d}.
The symmetry properties of the fermion operators 
under translation, rotation, and 
modified reflection are readily obtained by inspecting Eq.~(\ref{JW})
(throughout, we ignore a possible global $U(1)$ phase). 
In each case, the transformation of the Chern-Simons field is obtained 
from Eq.~(\ref{Adef}).
 
\begin{table} 
\caption{\label{tab:d} Symmetry transformation properties of the fields 
in the fermionized representation.  The transformation properties
of $a, e$ are the same as in Table~\ref{tab:dual}.
In the ${\mathcal C}$ column, $j=1,2$ refers to the sublattice 
index on the honeycomb lattice (site index in the unit cell pictured 
in Fig.~\ref{dual}).  The additional column ${\mathcal T}_{\rm ferm}$
corresponds to the na\"{\i}ve time reversal for the lattice fermions and
is \emph{not} a symmetry of the vortex Hamiltonian.} 
\begin{ruledtabular} 
\begin{tabular}{c | c | c | c | c || c} 
  & $T_{\delta {\bf r}}$, $R_{\pi/3}$  
  & $\tilde{\mathcal R}_x$ 
  & ${\mathcal C}$ 
  & ${\mathcal T}$ 
  & ${\mathcal T_{\rm ferm}}$ \\   
  \hline 
  $a\to$
  & $a$  
  & $-a$ 
  & $-a$ 
  & $-a$ 
  & $-a$ \\ 
  \hline 
  $e \to$ 
  &$e$ 
  & $e$ 
  & $-e$ 
  & $e$ 
  & $e$ \\ 
  \hline 
  $d_\bfx \to$
  & $d$ 
  & $d$
  & $(-1)^j d^\dagger$ 
  & $d_\bfx e^{-2 i \sum_{\bfx' \neq \bfx} \arg(\bfx, \bfx') N_{\bfx'} }$
  & $d$
  \\ 
  \hline 
  $A \to $
  & $A$ 
  & $-A$  
  & $-A$ 
  & $A$ \\ 
\end{tabular} 
\end{ruledtabular} 
\end{table}

Charge conjugation and time reversal require some explanation. 
From Eq.~(\ref{JW}), we have for particle-hole  
\begin{equation} 
{\cal C}: d_\bfx \to d_\bfx^\dagger  
\exp[i \sum_{\bfx' \neq \bfx} \arg(\bfx, \bfx') ] 
= d_\bfx^\dagger \exp[i \gamma_\bfx] ~. 
\end{equation} 
The phase $\gamma_\bfx$ is constant in time but depends on the 
lattice site.  For two nearest-neighbor sites on the honeycomb 
lattice we have 
\begin{eqnarray*} 
\gamma_{\bfx_1} - \gamma_{\bfx_2} &=& \arg(\bfx_1, \bfx_2) 
- \arg(\bfx_2, \bfx_1) \\ 
&& + \sum_{\bfx' \neq \bfx_1,\bfx_2}  
     [\arg(\bfx_1, \bfx') - \arg(\bfx_2, \bfx')] \\ 
&=& \pi - 2 A_{\bfx_1 \bfx_2}[N_{\bfx'}\equiv 1/2] ~. 
\end{eqnarray*} 
With our convention in Eq.~(\ref{Aconstraint}) where the Chern-Simons 
field fluctuates 
around zero, we conclude that $e^{i\gamma_\bfx}$ changes 
sign going between nearest neighbors.  The resulting 
$e^{i\gamma_\bfx}$ is written as $(-1)^j$ in Table \ref{tab:d}, 
where $j=1,2$ refers to the sublattice index of site $\bfx$. 

Finally, time reversal acts as 
\begin{equation} 
{\cal T}: d_\bfx \to d_\bfx  
\exp[- 2 i \sum_{\bfx' \neq \bfx} \arg(\bfx, \bfx') N_{\bfx'} ] .
\label{Tfermop} 
\end{equation} 
This is a formally exact implementation of the symmetry, but the phases 
multiplying the fermion operators now depend on the positions of all the 
vortices.  This nonlocal transformation represents a serious difficulty 
once we derive a low-energy continuum theory in the next subsection.  
In particular, it will not be possible to correctly represent the 
time-reversal transformation ${\cal T}$ using only the continuum fields.
Since our treatment below will involve crucial assumptions regarding
this issue, it is useful to also give a formulation of  
time-reversal symmetry in first-quantized language.  
Focusing on the particle degrees of freedom, the Jordan-Wigner 
transformation expresses the wavefunction for the bosonic vortices as 
\begin{equation} 
  \Psi_{\rm boson}(\bfx_1, \bfx_2, \dots) 
  = \prod_{i<j} e^{-i \arg(\bfx_i, \bfx_j)} \;\; 
  \Psi_{\rm ferm}(\bfx_1, \bfx_2, \dots) ~.
  \label{PsiJW} 
\end{equation} 
When the Chern-Simons phase factor is not affected by a 
symmetry transformation, 
the properties of the bosonic and fermionic wavefunctions coincide 
under the transformation.  However, time reversal sends $i \to -i$, 
hence the requirement that the bosonic wavefunction is real 
implies that
\begin{equation} 
\Psi^*_{\rm ferm}(\bfx_1, \bfx_2, \dots) =  
\prod_{i<j} e^{-2 i \arg(\bfx_i, \bfx_j)} \;\;  
\Psi_{\rm ferm}(\bfx_1, \bfx_2, \dots) ~.
\label{PsifermT}
\end{equation}
Thus, the time-reversal invariance of the bosonic wave function 
is a highly nontrivial condition for the fermionic wavefunction.

For this reason, it is convenient to define a {\it modified}
time-reversal transformation, which corresponds to na\"{\i}ve time reversal 
for the lattice fermions:
\begin{equation}
{\cal T}_{\rm ferm}: \quad d_{\bf x} \to d_{\bf x};
\quad a \to -a;  \quad e \to e;  \quad i \to -i.
\label{Tferm}
\end{equation}
In first-quantized language ${\cal T}_{\rm ferm}$ corresponds to 
complex conjugation of the wavefunction for the fermionized vortices, 
$\Psi_{\rm ferm} \to \Psi^*_{\rm ferm}$.  It is important to emphasize 
that this transformation is {\it not} a symmetry of the fermionic 
Hamiltonian, since under ${\cal T}_{\rm ferm}$ the field $a$ changes
sign while the Chern-Simons field $A$ from Eq.~(\ref{Adef}) remains 
unchanged.
 
However, as we shall argue further below, since the vortices interact
logarithmically, their density fluctuations are greatly 
suppressed, and it is plausible that the phase factors in 
Eqs.~(\ref{Tfermop}) and (\ref{PsifermT}) might be essentially
the same for all vortex configurations that carry substantial weight.  
For an extreme example, if the vortices form a perfect charge-ordered 
state, then the Chern-Simons phase factor is constant.
We conjecture below that this is also the case at the critical points 
separating the spin-ordered states from the spin paramagnet.
In any event, in situations where the Chern-Simons phase factors are roughly 
constant for all important configurations, the physical and modified
time-reversal transformations become essentially identical.  
It is then legitimate to require that the theory be 
invariant under ${\cal T}_{\rm ferm} $.  This will be useful when we 
describe a conjectured low-energy continuum fermionic theory for 
criticality in the vortex system, since ${\cal T}_{\rm ferm} $ acts 
in a simple, local way on the continuum fermion fields.

\subsection{Na\"{\i}ve Continuum Theory} 
\label{subsec:Continuum}
To arrive at a low-energy continuum theory, we first consider a  
``flux-smeared'' mean-field state with $A_{\bfx \bfx'} = 0$.   
We will also ignore fluctuations in $a_{\bfx \bfx'}$ for the moment,  
taking $a_{\bfx \bfx'} = 0$.
We are then left with half-filled fermions hopping on the  
honeycomb lattice with no fluxes. 
 
Diagonalizing the hopping Hamiltonian in momentum space using the 
two-site unit cell shown in Fig.~\ref{dual}, one finds that there are 
two Dirac points at momenta $\pm{\bf Q}$, where  ${\bf Q} = (4\pi/3,0)$.
(Note that these are the same wavevectors found for the low-energy 
spin-1 excitations in the continuum analysis of the original spin model 
in Sec.~\ref{subsec:DirectLGW}.) 
Focusing only on low-energy excitations in the vicinity of these  
momenta, the fermion operators can be expanded around the Dirac points.
Denoting the lattice fermion field on the two sublattices as 
$d_{{\bf x}\alpha}$ with $\alpha = 1,2$ the sublattice label, we write 
\begin{equation} 
  d_{{\bf x}\alpha}  \approx  
  \psi_{R\alpha}  e^{i{\bf Q}\cdot \bfx}  
  + i\sigma_{\alpha \beta}^y \psi_{L \beta} e^{-i {\bf Q}\cdot \bfx}, 
  \label{psi}
\end{equation} 
where ${\bf x}$ continues to label the real-space position of the 
honeycomb lattice site and $\sigma^y$ is a Pauli matrix.
The fields $\psi_{R/L}$ are two-component spinors that vary 
slowly on the lattice scale.  Using this expansion 
we obtain for the free-fermion part of the continuum Hamiltonian
\begin{equation} 
  H_f^{(0)} = \int d\bfx \; \psi_a^\dagger \, v \, 
  (p_x \sigma^x + p_y \sigma^y) \psi_a ~, 
  \label{contH}
\end{equation} 
where ${\bf p}=-i\bm{\nabla}$ is the momentum operator, 
an implicit summation over the flavor index $a \in [R,L]$ is understood,
and we have suppressed the summation over the spinor indices 
$\alpha,\beta$.
From now on, we absorb the nodal velocity $v$ in the scaling of the 
coordinates. 
 
It is convenient to work in the Euclidean path integral description. 
The free-fermion Lagrangian density is written in the form 
\begin{eqnarray} 
  {\mathcal L}_f^{(0)} &=& \barpsi_a \gamma^\mu \partial_\mu \psi_a, 
  \\ 
  \barpsi_{R/L} &\equiv& \psi_{R/L}^\dagger\gamma^0, 
\end{eqnarray} 
with implicit sums over the flavor index $a \in[R,L]$ and  
space-time index $\mu \in[0,1,2]$, defined so that  
$\partial_{0,1,2} \equiv \partial_{\tau,x,y}$.   
The $2\times 2$ Dirac matrices $\gamma^\mu$ are given by  
$\gamma^0 = \sigma^z$, $\gamma^1 = \sigma^y$, $\gamma^2 = -\sigma^x$  
and satisfy the usual algebra 
$\{\gamma^\mu,\gamma^\nu\} = 2 \delta^{\mu\nu}$.  
These matrices act within each two-component field so that  
$(\gamma^\mu \psi_a)_\alpha \equiv \gamma^\mu_{\alpha \beta}
\psi_{a \beta}$.   
We will also find it useful to define Pauli matrices $\tau^{x,y,z}$ 
that act on the $R/L$ flavor indices, i.e.,  
$(\tau^k \psi)_{a \alpha} \equiv \tau^k_{ab}\psi_{b \alpha }$.   
 
Including the gauge-field fluctuations and the short-range fermion  
interactions, we arrive at a Lagrangian of the form 
\begin{eqnarray} 
  {\mathcal L} &=& 
  \barpsi_a \gamma^\mu (\partial_\mu - i a_\mu - i A_\mu )\psi_a 
  \nonumber \\ 
  &+& \frac{1}{2 e^2} (\epsilon_{\mu\nu\lambda}\partial_\nu a_\lambda)^2 
  + \frac{i}{4\pi} \epsilon_{\mu\nu\lambda}A_\mu \partial_\nu A_\lambda 
  + {\mathcal L}_{\rm 4f}. 
  \label{L} 
\end{eqnarray} 
Here $\epsilon_{\mu\nu\lambda}$ is the anti-symmetric tensor, and for 
simplicity we wrote a space-time isotropic form for the Maxwell action 
of the gauge field $a_\mu$.  We also used the standard $(2+1)$-dimensional form for 
the Chern-Simons action that ensures the attachment of $2\pi$ flux 
to the fermions, restoring the bosonic exchange statistics of the 
vortices.  In the absence of the four-fermion terms ${\mathcal L}_{4f}$, there is a global $SU(2)$ flavor symmetry, whose action on the fermion fields is generated by the $\tau^k$ Pauli matrices.
 
The four-fermion interaction terms can be written as
\begin{equation}
{\mathcal L}_{4f} = \lambda_0 J_0^2 + \lambda_z J_z^2 
   + \lambda_\perp (J_x^2 + J_y^2)  
   + \lambda^\prime_0 (\barpsi_a \gamma^0\psi_a)^2  , 
\label{L4f} 
\end{equation}
where we have defined a flavor $SU(2)$ vector of fermion bilinears,
\begin{equation}
\bm{J} \equiv \barpsi_a {\bm\tau}_{ab} \psi_b 
       = \psi^\dagger_a \sigma^z {\bm\tau}_{ab} \psi_b ~, 
\end{equation}
and a ``mass term,"
\begin{equation}
J_0 \equiv \barpsi_a \psi_a = \psi^\dagger_a \sigma^z \psi_a ~.
\end{equation}
The four-fermion terms arise from vortex density-density 
interactions and other short-range interaction processes.  
For example, the $\lambda_0^\prime$ term roughly represents an overall 
vortex repulsion, while the $\lambda_z$ term represents the difference in
repulsion between vortices on the same sublattice and opposite 
sublattices of the honeycomb. 
${\mathcal L}_{4f}$ contains all independent four-fermion terms 
that can arise from the microscopic short-range fermion interactions 
(including possible short-range pieces mediated by the gauge fields).

The above constitutes the na\"{\i}ve continuum limit obtained by inserting 
the slow-field expansion Eq.~(\ref{psi}) into the microscopic 
Hamiltonian and assuming small fluctuations in the gauge field $a$ and 
the Chern-Simons field $A$.  We now discuss the symmetries of the 
continuum formulation.  Table~\ref{tab:psi} shows the transformation 
properties of the continuum fermion fields deduced from the lattice 
fermion transformations in Table~\ref{tab:d}.  Also shown are the 
symmetry transformation properties of the fermion bilinears $\bm{J}$ 
and $J_0$.  Missing from the table is the original time-reversal 
invariance, which we do not know how to realize in the continuum.  
In its place we have included the modified time reversal 
${\cal T}_{\rm ferm}$, defined in Eq.~(\ref{Tferm}), which corresponds 
to the na\"{\i}ve time-reversal transformation for fermions hopping on the 
honeycomb lattice.  Remarkably, the transformation properties of the 
flavor $SU(2)$ vector $\bm{J}$ of fermionic bilinears are identical 
to the transformation properties of the bosonic flavor $SU(2)$ vector 
defined in Sec.~\ref{subsec:DirectLGW}, 
$\bm{K} = \Phi^*_a \bm{\tau}_{ab} \Phi_b$.  
Surprisingly, the fermionic bilinear $\bm{J}$ transforms under the 
{\it modified} time-reversal transformation ${\cal T}_{\rm ferm}$
in precisely the way that the bosonic bilinear $\bm{K}$ transforms 
under the physical time reversal ${\cal T}$.

\begin{table}
\caption{\label{tab:psi} Transformation properties of the continuum 
fermion fields $\psi_{R/L}$ and the fermion bilinears $\bm{J}$, $J_0$
under the microscopic symmetries.  Here $J_\pm = J_x \pm i J_y$. 
The spatial coordinates, which should transform appropriately under the 
lattice symmetries, have been suppressed for conciseness.  
The last column {\em does not} correspond to the physical time-reversal 
symmetry, which we cannot realize in a local manner in the continuum 
theory.  Rather, it corresponds to the na\"{\i}ve time reversal in
Eq.~(\ref{Tferm}) for fermions hopping on the honeycomb lattice.} 
\begin{ruledtabular} 
\begin{tabular}{c | c | c | c | c || c} 
  & $T_{\bf \delta r}$ & $R_{\pi/3}$ & $\tilde {\cal R}_x$ 
  & ${\mathcal C}$ & ${\cal T}_{\rm ferm}$ \\ 
  \hline 
  $\psi \to$  
  & $e^{i{\bf Q\cdot \delta r}\tau^z} \psi$ 
  & $i \tau^x e^{-i \pi \sigma^z/6} \psi$ 
  & $\psi$
  & $\tau^x \sigma^x [\psi^\dagger]^t$  
  & $i\tau^y \, i\sigma^y \psi$ \\
  \hline 
  $J_z \to$  
  & $J_z$ 
  & $-J_z$
  & $J_z$  
  & $-J_z$  
  & $J_z$ \\
  \hline 
  $J_\pm \to$  
  & $e^{\mp 2 i {\bf Q\cdot \delta r}} J_\pm $ 
  & $J_\mp$
  & $J_\pm$  
  & $J_\pm$  
  & $J_\mp$ \\ 
\hline 
$J_0 \to$  
  & $J_0 $ 
  & $J_0$
  & $J_0$  
  & $J_0$  
  & $-J_0$
\end{tabular} 
\end{ruledtabular} 
\end{table}

One can verify that the first four symmetries in 
Table \ref{tab:psi} (translation, rotation, modified reflection, and 
particle-hole) preclude all quartic fermion terms from the Lagrangian except
the four terms exhibited in Eq.~(\ref{L4f}).  
Moreover, these symmetries prohibit all fermionic 
bilinears in the Lagrangian except a mass term of the form 
$M J_0 = M \barpsi \psi$.  This merits some discussion.  In situations 
where it is legitimate to replace the time-reversal transformation
by the modified symmetry ${\cal T}_{\rm ferm}$, we can use this modified 
symmetry to preclude such a mass term.  However, generally this 
might not be possible.  Indeed, as we shall see in the next subsection, 
a large mass term of this form when added to the Lagrangian places the 
system in the spin paramagnetic phase, {\it provided} the mass $M$
has a specific sign relative to the Chern-Simons term.  Since we know that 
the spin paramagnet does not break time-reversal symmetry ${\cal T}$, 
it is clear that at least in this phase it is {\it not} legitimate to 
replace ${\cal T}$ with ${\cal T}_{\rm ferm}$.
The reasons for this will become clear in the next subsection where we
describe how the spin paramagnet and the spin ordered phases can be 
correctly described using the fermionized-vortex formulation.
The discussion of whether or not it is legitimate to replace
${\cal T}$ with ${\cal T}_{\rm ferm}$ at the critical points separating 
these phases will be deferred until Sec.\ \ref{subsec:FermCrit}.

\subsection{Phases in the Fermionic Representation}
\label{subsec:FermPhases}
We now discuss how to recover the phases of the original spin model 
using the fermionic vortex fields.  This extends our earlier discussion 
using the bosonic vortices in Sec.~\ref{subsec:BosonicPhases}. 
The paramagnetic phase of the original spin model is the bosonic 
vortex superfluid (Higgs) phase.  In terms of the fermionic vortex 
fields, the paramagnet is obtained as an integer quantum Hall state, 
which in the continuum description corresponds to the presence of a 
mass term $M \barpsi \psi$.  Both fermion fields $\psi_R$ and $\psi_L$ 
then have the same mass $M$ with the same sign. 
Let us see that we indeed recover the correct description of the 
original spin paramagnet.  Integrating out the massive fermions 
induces a Chern-Simons term for the sum field $(a+A)_\mu$, so the Lagrangian 
for the fields $a_\mu$ and $A_\mu$ becomes 
\begin{eqnarray} 
  {\cal L}_{a, A} &=&  
  \frac{1}{2 e^2} (\epsilon_{\mu\nu\lambda}\partial_\nu a_\lambda)^2 + 
  \frac{i}{4\pi} \epsilon_{\mu\nu\lambda} A_\mu \partial_\nu A_\lambda \\ 
  && + \frac{i \text{sign}(M)}{4\pi}  
  \epsilon_{\mu\nu\lambda}(a+A)_\mu \partial_\nu (a+A)_\lambda ~.
\end{eqnarray}
We now stipulate that the sign of the mass $M$ is taken to 
cancel the original Chern-Simons term for the field $A$.  
One can then verify
that the spectrum corresponding to the above Lagrangian is gapped.
For example, upon integrating out $A$ the gauge field $a$ obtains a mass.
This is as expected, since there are no gapless excitations
in this phase.  Gapped quasiparticles of the paramagnetic phase 
of the spin model are described as follows. 
Consider acting with the fermion field $\psi_{R/L}^\dagger$ on the  
ground state.  This is essentially the same as inserting a  
{\em vortex in the vortex field $\theta$}.  The added fermion
couples to $a+A$, and by examining ${\cal L}_{a,A}$ we find that
it binds $\nabla \!\times\! a = -2\pi$ flux to the fermion.  Thus, 
the fermion is
turned into a localized bosonic excitation, which is the familiar
screened vortex in the vortex field $\theta$ 
(the sign of the flux is of course consistent with our minimal 
coupling convention ${\bm \nabla}\theta + {\bm a}$).
In terms of the original spin model, this bosonic excitation
carries spin 1 and is precisely the magnon of the paramagnet.

It is worth remarking about the role of time-reversal symmetry and the 
modified time-reversal transformation ${\cal T}_{\rm ferm}$ of 
Eq.~(\ref{Tferm}) in the spin paramagnetic phase.  Since the spin 
paramagnet corresponds to an integer quantum Hall state for the 
fermions, it is clear that ${\cal T}_{\rm ferm}$ will {\it not} be 
respected in this phase.  
This is consistent with Table~\ref{tab:psi}, which shows that
the mass term $\barpsi \psi$ is odd under ${\cal T}_{\rm ferm}$.  
On the other hand, the phase factors in the fermionic integer quantum Hall 
wavefunction 
will conspire to cancel the Chern-Simons phase factors in 
Eq.~(\ref{PsifermT}) leading to a wavefunction for the bosonic vortices 
which is real---consistent with physical time-reversal invariance.

Consider now the magnetically ordered phases of the original spin model. 
These correspond to vortex insulators and are obtained in the  
fermionic theory as a result of spontaneously generating a fermion mass  
of the form
\begin{equation}
\la \bm{J} \ra = \la \barpsi \bm{\tau} \psi \ra \neq 0 ~.
\label{Jorder}
\end{equation}
In the presence of such mass terms we have two massive Dirac 
fermion fields with opposite-sign masses, and integrating out the fermions 
produces only a generic Maxwell term for the sum field $(a+A)_\mu$.
The gapless photon mode of the gauge field $a$ then corresponds to the  
spin wave of the magnetically-ordered phase. 
Acting with a fermion creation operator also binds $2\pi$ flux of the  
Chern-Simons field $A$ and turns the fermion back into the original  
bosonic vortex.  Because of the gapless gauge field $a$ such an  
isolated vortex costs a logarithmically-large energy as expected 
in the spin-ordered phase. 
 
\begin{figure} 
\begin{center} 
{\resizebox{8cm}{!}{\includegraphics{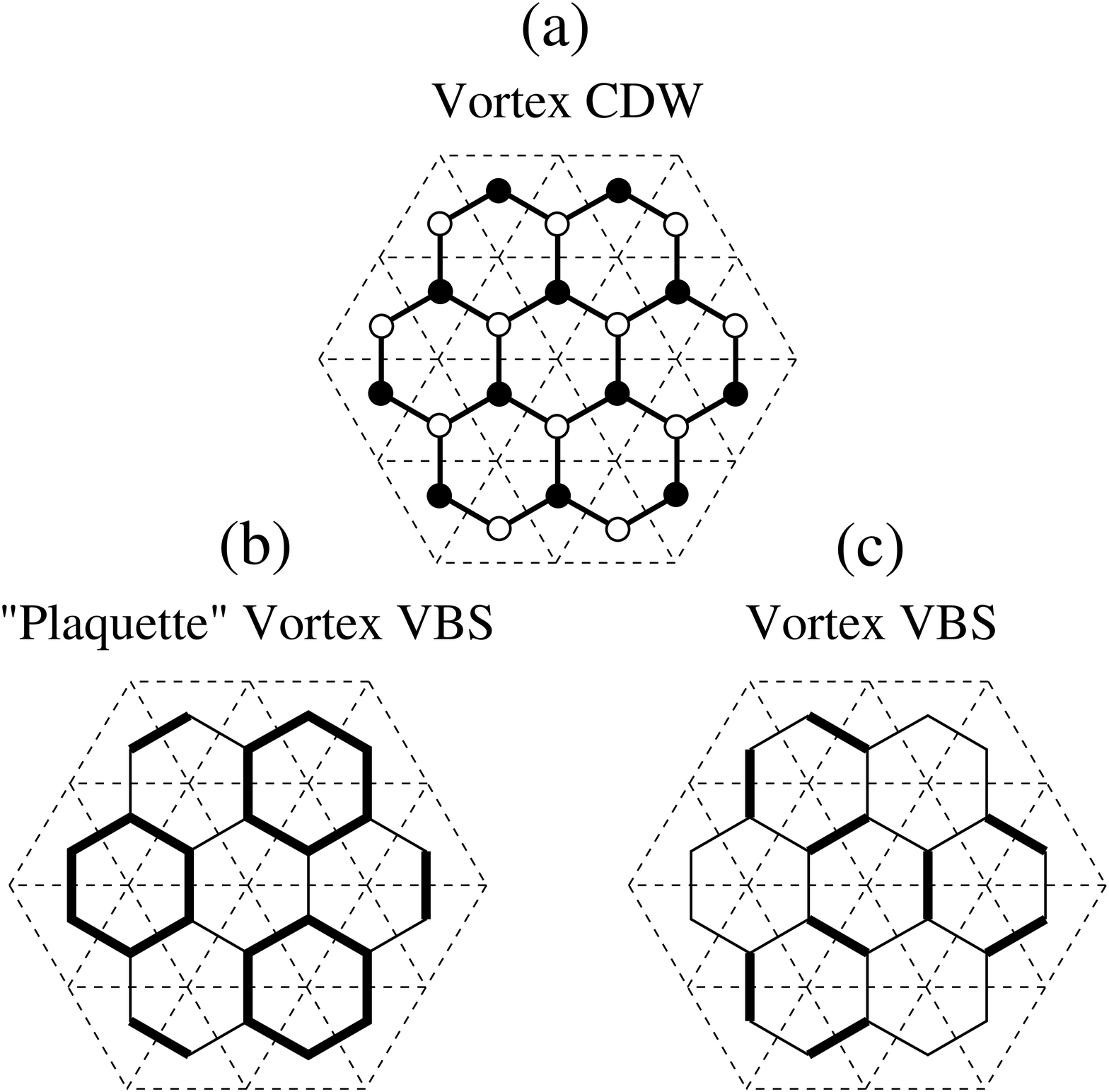}}} 
\end{center} 
\caption{Vortex insulators that correspond to the spin-ordered states
of Fig.~\ref{SpinSts}.
(a) Staggered vortex ``charge density wave'' (CDW) 
state, corresponding to the coplanar 
$\sqrt{3}\times\sqrt{3}$ state in the original spin variables.   
Vortices preferentially occupy one of the two sublattices.
(b) ``Plaquette'' vortex valence bond solid (VBS), corresponding to the 
dice-lattice collinear antiferromagnetic spin order.  Vortex density is spread
over the bold hexagons.
(c) Vortex VBS, corresponding to the hexagonal
collinear antiferromagnetic spin order.  Vortices preferentially
occupy the lattice of bold links.} 
  \label{VortSts} 
\end{figure}

Spontaneous mass generation is driven by interaction 
terms in the Hamiltonian such as the $\lambda_z$ and $\lambda_\perp$ 
interactions.  The details of the magnetic order are determined by the 
specific mass term that is generated.  For example, the mass term 
$m_z J_z $ in the Lagrangian corresponds to the vortex ``charge density
wave'' (CDW) state wherein vortices preferentially occupy 
one sublattice of the honeycomb lattice as shown in Fig.~\ref{VortSts}a.  
Indeed, from Table~\ref{tab:psi}, this mass term is odd under the 
$\pi/3$ rotation and particle-hole symmetries, and can be identified 
with a staggered chemical potential for the lattice fermions that 
selects one of the charge-ordered states over the other.  Therefore such 
spontaneously-generated mass gives rise to the translation symmetry 
breaking in the vortex system that produces the CDW state. 
In the original spin model, this corresponds to the coplanar spin
state of Fig.\ \ref{SpinSts}a.

On the other hand, the mass term $m_x J_x + m_y J_y$ corresponds 
to the vortex valence bond solid (VBS) state.  The specific pattern is not 
resolved with only the four-fermion interactions, but we expect that 
because of the underlying lattice there are higher-order terms in the 
action that pin the direction in the $(J_x, J_y)$ plane so that 
$\la J_+ \ra^3$ is either $+1$ or $-1$.  The corresponding bond orders
are deduced by interpreting the spontaneously-generated mass term 
$|m| (e^{i\alpha} J_+ + e^{-i\alpha} J_- )/2$ as inducing a modulated 
vortex hopping amplitude
$t_{\bfx, \bfx'} = 1 + |m| \cos[{\bf Q} \cdot (\bfx + \bfx') + \alpha]/2$.
For $e^{i 3\alpha} = -1$ the stronger bonds form hexagons and produce
the vortex ``plaquette'' VBS shown in Fig.~\ref{VortSts}b,
which in the original spin model corresponds to the dice-lattice collinear
spin order of Fig.\ \ref{SpinSts}b.  
On the other hand, for $e^{i 3\alpha} = 1$ the strong
bonds form the lattice shown in Fig.~\ref{VortSts}c; 
the resulting vortex VBS corresponds to the 
hexagonal collinear spin state in Fig.\ \ref{SpinSts}c.

It is worth recalling from Sec.~\ref{subsec:DirectLGW} that in the 
Landau theory of the spin model, the chirality order parameter 
($\kappa$) that develops in the coplanar state and the bond energy wave 
(complex) order parameter ($B$) that is non-vanishing in the collinear 
spin states are both expressible in terms of the $SU(2)$ vector $\bm{K}$ 
defined in Eq.~(\ref{Kvec}):
\begin{equation}
\kappa \sim \la K_z \ra; \quad 
B \sim \la K_x + i K_y \ra ~. 
\end{equation}
Thus, the magnetically-ordered states are characterized by
\begin{equation}
\la \bm{K} \ra = \la \Phi^* \bm{\tau} \Phi \ra \ne 0  ,
\end{equation}
with coplanar order corresponding to ordering $K_z$ and collinear order
corresponding to an ordering in the $(K_x, K_y)$ plane.  
This is directly analogous to the ordering patterns of the $SU(2)$ vector 
$\bm{J} = \barpsi \bm{\tau} \psi$ as discussed above.  
 
The fermionic formulation of the dual vortex theory thus allows us to 
correctly describe the phases of the original spin model.  
From the point of view of the bosonic vortex system at half-filling, 
this is rather nontrivial.  For example, the fermionic formulation 
correctly captures the two low-energy spin-1 excitations with 
wavevectors $\pm {\bf Q}$ in the spin paramagnet phase, even though 
these wavevectors are not apparent in terms of the bosonic vortices.
Moreover, the magnetically ordered phases are accessed in a unified 
manner within the fermionic formulation, via a spontaneous mass 
generation driven by the vortex interactions which produces either 
the vortex CDW or one of the VBS states, thereby ``releasing'' the
dual photon.  While the vortex CDW is perhaps natural in the bosonic 
vortex theory given the strong repulsive vortex interactions and the 
specific charge ordering, the VBS phases are nontrivial for 
the vortex bosons at half-filling.  Indeed, a common approach for 
analyzing such VBS states of bosons is to study their dual 
formulation,\cite{Lannert}
which in the present context corresponds to our analysis of the original 
spin model.  Our alternative route via the fermionization achieves  
this due to the fact that it in some sense combines the direct and  
dual perspectives.

\subsection{Criticality in the Fermionic Theory?}
\label{subsec:FermCrit}
Encouraged by these successes, we now embark on a study of the  
transitions between the phases using the fermionic formulation. 
As a guide we use the anticipation that the original spin model 
has continuous 
transitions.  We have confidence for the presence of the $O(4)$ 
fixed point and its likely appearance as a multicritical point, 
and also for the continuous decoupled $U(1)\times U(1)$ 
transition between the  
paramagnet and the collinearly-ordered phase.  Moreover, the  
paramagnet-to-coplanar ordering transition may also be continuous. 
We want to now see if we can access any of these critical points
within the Dirac theory of fermionized vortices.
 
At criticality, we expect the vortices to be massless.  We 
also expect the original spins to be critical, which corresponds to 
the dual gauge field $a$ being critical. 
To proceed we first perform the following trick in the continuum 
action.  The fermion current couples to the combination 
$\tilde{a}_\mu = a_\mu + A_\mu$, and it is instructive to retain only this  
field in the path integral by performing the Gaussian integration
over the field $A$.  The resulting na\"{\i}ve continuum Lagrangian has the 
form 
\begin{eqnarray} 
 {\mathcal L} &=& \barpsi_a  
                  \gamma^\mu (\partial_\mu - i \tilde{a}_\mu )\psi_a 
  + {\cal L}_{4f} 
  \nonumber \\ 
  &+& \frac{1}{2 e^2} (\nabla \times \tilde{a})^2
  + i \frac{\pi}{e^4} (\nabla \times \tilde{a}) \cdot  
          (\nabla \times \nabla \times \tilde{a}) ~.
\label{Lfinal}
\end{eqnarray} 
Since $\tilde{a}$ should be massless at criticality, it seems likely 
that the last term above (which can be viewed as a higher-derivative 
Chern-Simons term) is irrelevant compared with the Maxwell term at long 
wavelengths.  Indeed, ignoring the effects of this term on the field 
$\tilde{a}_\mu$ gives a correct description in the vortex insulator 
phases, where the fluctuations of $\tilde{a}_\mu$ are essentially the 
same as $a_\mu$ and represent gapless spin waves of the original 
spin model. 
As a putative description of criticality in the original spin model,
we henceforth drop this term altogether.  The theory is then equivalent 
to quantum electrodynamics in (2+1) dimensions, 
\begin{eqnarray} 
{\mathcal L}_{QED3} =  
\barpsi_a \gamma^\mu (\partial_\mu - i \tilde{a}_\mu) \psi_a 
+ \frac{1}{2 e^2} (\nabla \times \tilde{a})^2
+ {\mathcal L}_{4f} ~. 
\end{eqnarray} 
We conjecture that this QED3 theory with two Dirac fermions contains 
a description of the critical properties of the original spin 
model. 

Words of caution are in order here.  Dropping the higher-derivative 
Chern-Simons term changes the symmetry of the 
continuum Lagrangian, which is now 
invariant under the modified time-reversal symmetry 
${\cal T}_{\rm ferm}$, as can be seen from Table~\ref{tab:psi}.  
Apparently, neglecting the higher-derivative Chern-Simons 
term is tantamount to 
replacing the physical time-reversal symmetry transformation
by the modified one.  This seems reasonable on physical 
grounds since the vortices are strongly-interacting, and 
power-counting in the Lagrangian lends further mathematical 
support to the validity of this approximation.  
Once we have replaced ${\cal T}$ by ${\cal T}_{\rm ferm}$,
both the fermion mass term $M \barpsi\psi$ and the higher-derivative 
Chern-Simons term are symmetry precluded.  This seems consistent 
since we are looking for a fixed-point theory with massless fields.

It is worth emphasizing that the above QED3 Lagrangian is the proper 
continuum theory for fermions hopping on the honeycomb lattice  
coupled to a noncompact gauge field (no Chern-Simons field).
Invariance under the modified time-reversal symmetry 
${\mathcal T}_{\rm ferm}$ follows 
provided the fermionic hopping Hamiltonian is real.  
Our proposal is that such a critical QED3 theory faithfully describes the 
continuous transitions of our bosonic vortices at half-filling.
Again, the strong logarithmic interaction between vortices greatly 
suppresses their density fluctuations, and one expects that as a
result the vortex 
statistics might not be so important.  The trick that eliminated the 
Chern-Simons gauge field $A$ leaving only the 
higher-derivative Chern-Simons-like term for 
$\tilde{a}$ can be viewed as a more formal argument for this.
If we are allowed to drop such higher-derivative terms, then we can 
essentially eliminate the statistical Chern-Simons field by absorbing it 
into the already-present gauge field $a$, which is precisely the field 
that mediates the vortex repulsion.
 
Assuming that QED3 is appropriate for describing criticality in the 
spin model, it is not at first clear whether or not this requires  
a further fine-tuning of the mass term $M \barpsi\psi$ to zero in the
continuum theory.  However, once we assume that it is legitimate
to replace ${\cal T}$ by ${\cal T}_{\rm ferm}$ at criticality, then this  
fermion mass term is symmetry-precluded, and we are not allowed to add it
to the critical Lagrangian as a perturbing field.

In the next section we will analyze the QED3 Lagrangian \emph{per se},
focusing on its potential to describe criticality in the spin model.

\section{Analysis of QED3}
\label{sec:QED}
The QED3 theory with two Dirac fermions as realized on the half-filled
honeycomb lattice is a difficult problem with its own phase diagram.
As we will argue below, it is likely that the lattice model generically
ends up in a phase with a spontaneously-generated fermion mass.
In this case, the continuum Lagrangian ${\cal L}_{QED3}$ with massless
fields potentially applies only to critical points of the lattice model.

\subsection{Critical Theory}
\label{subsec:QEDCrit} 
In addition to the discrete symmetries tabulated in Table~\ref{tab:psi},
the Lagrangian ${\cal L}_{QED3}$ has a number of continuous global 
symmetries.  The full Lagrangian is invariant under the (dual) global 
$U(1)$ symmetry, $\psi_a \to e^{i\alpha}\psi_a$.  Associated with this 
symmetry is a conserved vortex current $G^\mu = \barpsi \gamma^\mu \psi$, 
which satisfies $\partial_\mu G^\mu = 0$.  Here $G^0$ is simply the 
vortex density.  In the absence of the four-fermion interaction terms, 
the Lagrangian also enjoys a global flavor $SU(2)$ symmetry, being 
invariant under
\begin{equation}
  \psi_a \to U_{ab} \psi_b; \quad\quad 
  \barpsi_a \to \barpsi_b U^\dagger_{ba}  ,
  \label{SU2}
\end{equation}
for arbitrary $SU(2)$ rotation, $U^\dagger = U^{-1}$.  The associated 
$SU(2)$ conserved currents are given by 
$\bm{j}^\mu = \barpsi \gamma^\mu \bm{\tau} \psi$, and satisfy 
$\partial_\mu \bm{j}^\mu =0$.
The remaining fermionic bilinears $J_0 = \barpsi\psi$ and
$\bm{J} = \barpsi \bm\tau \psi$ are not parts of any conserved
current; $J_0$ is a scalar while $\bm{J}$ rotate as a vector under the 
flavor $SU(2)$.
In the special case $\lambda_z = \lambda_{\perp}$, the four-fermion 
interaction terms are also invariant under $SU(2)$ flavor rotations, 
but more generally are not.

Due to the gauge interactions QED3 is a strongly-interacting field 
theory\cite{Appelquist}.  Specifically, expanding about the free-field limit with 
$e^2=0$ and ${\mathcal L}_{4f}=0$, the continuum fermion 
fields have scaling dimension 
$\Delta_\psi =1$ so that the four-fermion interactions are irrelevant.
However, gauge invariance dictates that $\tilde a$ has scaling
dimension $\Delta_{\tilde a}=1$, 
implying that $e^2$ is relevant and grows in the infrared.  

To seek a controlled limit one is forced to generalize the model
in some way.  Perhaps the simplest approach is to introduce $N$ copies of the 
fermion fields, $\psi_{a\alpha} \to \psi_{ja\alpha}$ with $j=1,\dots,N$, 
each of which is minimally coupled to the same gauge field, 
and to then study the model in the large-$N$ 
limit.\cite{Appelquist, Rantner, Vafek, Franz, Herbut, Hermele}
(Note that the theory with two fermion flavors corresponds to 
$N = 1$, and the flavor symmetry of the theory thus 
generalized is $SU(2 N)$.)  
Upon integrating out the fermions and expanding to quadratic order in 
the gauge field one obtains an effective gauge action of the form,
\begin{equation}
S_a = \frac{1}{2g} \int \frac{d^3\bm{q}}{(2 \pi)^3} \;
|\bm{q}|\, \left(\delta_{\mu\nu} - \frac{q_\mu q_\nu}{|\bm{q}|^2}\right)
\tilde{a}_\mu(-\bm{q}) \tilde{a}_\nu(\bm{q})~,
\label{gaugescreened}
\end{equation}
with a small coupling constant $g = 8/N$.  
The gauge propagator proportional to $g/|\bm{q}|$ mediates a screened 
interaction between the fermions which falls off as $1/|\bm{r}|$ in real 
space and is much weaker than the bare logarithmic interaction.
At infinite $N$ the gauge fluctuations are completely 
suppressed, and, except for some subtleties that will not be important here,\cite{Hermele} the model reduces to free Dirac fermions.  
It is then possible\cite{Appelquist, Rantner, Vafek, Franz, Herbut, Hermele}
to perform a controlled analysis perturbative in 
inverse powers of $N$.  Specifically, one can compute the scaling 
dimension of various perturbations, such as the quartic fermion terms, 
order by order in $1/N$.  

To obtain the leading $1/N$ corrections it suffices to retain only the 
original two fermion fields $\psi_{R/L}$ that appear in 
${\cal L}_{QED3}$ and to replace the Maxwell term by the singular gauge 
interaction from Eq.~(\ref{gaugescreened}).  One can then perform a 
simple Wilsonian renormalization group analysis perturbative in the single 
coupling constant $g$.  After integrating out a shell of modes between a 
cutoff $\Lambda$ and $\Lambda/b$ with $b>1$, the fermion fields can be 
rescaled to keep the $\barpsi \gamma^\mu \partial_\mu \psi$ term 
unchanged.  Gauge invariance then automatically ensures that 
$\barpsi \gamma^\mu \tilde{a}_\mu \psi$ is also unchanged.  
Due to the singular momentum dependence in Eq.~(\ref{gaugescreened}), 
$1/g$ cannot pick up any diagrammatic contributions from the 
high-momenta mode integration.  With $\Delta_{\tilde a}=1$ assured by gauge 
invariance, rescaling will not modify $g$, and the theory describes a 
fixed line parameterized by the coupling $g$.  

With this simple Wilsonian renormalization group in hand, one can easily 
compute the scaling dimensions of the quartic fermion operators 
perturbatively in $g$.  We find that to leading order in $g$ the 
scaling dimension of the $\lambda_0^\prime$ term is unmodified, with 
$\Delta_{\lambda_0^\prime} = 4 + O(g^2)$.

The other three quartic fermion terms mix already at first-order in $g$.
Of the three renormalization group eigenoperators, we find two that are
singlets under the flavor $SU(2)$,
\begin{equation}
  \hat{Q}_0 =  \bm{J}^2 - J_0^2 ; \quad\quad  
  \hat{Q}^\prime_0 =  \bm{J}^2 + 3 J_0^2  ,
\label{Q0}
\end{equation}
and one that transforms as spin 2 under the flavor $SU(2)$,
\begin{equation}
  \hat{Q}_2 = \bm{J}^2 - 3J_z^2  .
\end{equation}
To first order in $g$ the respective scaling dimensions are
\begin{equation}
  \Delta_{Q_0} = 4 - 4 g/\pi^2  ; \quad 
  \Delta_{Q_0^\prime} = \Delta_{Q_2} = 4 + 4 g/(3\pi^2)  .
  \label{scalingdim}
\end{equation}

The above discussion is based on a particular generalization of 
the terms in ${\mathcal L}_{4f}$ to the 
$SU(2N)$-symmetric theory, where quartic terms contain only two fermion
flavors.  Another natural generalization proceeds by 
classifying all four-fermion terms in the $SU(2N)$-symmetric theory 
according to the irreducible representation of flavor and Lorentz 
symmetry under which they transform.  It is possible to establish 
a natural correspondence between four of these multiplets and the 
multiplets in the $SU(2)$ theory to which the terms in 
${\mathcal L}_{4 f}$ belong.  One can then calculate the scaling 
dimensions of the resulting terms order-by-order in $1/N$.  For 
the terms corresponding to $\hat{Q}_0$, $\hat{Q}_0'$ and 
$\hat{Q}_2$, we reproduce the same scaling dimensions above.  
On the other hand, the analog of the $\lambda'_0$ term has 
dimension $6 + {\cal O}(1/N)$, as can be seen by a calculation 
of its autocorrelation function at $N = \infty$.  We remark that 
both generalizations of this term strongly suggest it is an 
irrelevant perturbation.

At sufficiently large $N$ all of the quartic terms have scaling 
dimensions greater than $D=2+1$, and are thus irrelevant.  
The Lagrangian ${\cal L}_{QED3}$ is then critical and describes a 
conformally-invariant, strongly-interacting fixed point.  
Our hope is that this fixed point (at $N=1$) corresponds to one of the 
three critical points of the original spin model discussed in Sec.\ 
\ref{subsec:DirectCrit}.  We will try to identify which one in the
next subsection.  In Sec.\ \ref{subsec:QEDStabl}, we 
will return to the important issue of the 
stability of this fixed point in the physically-relevant $N=1$ case.

\subsection{Multicritical Point in the Spin Model?}
\label{subsec:Multicrit}
In the absence of the quartic terms, the critical theory described by 
${\cal L}_{QED3}$ has a global $SU(2)$ symmetry shown explicitly in 
Eq.~(\ref{SU2}).  Since the three fermionic bilinears comprising 
$\bm{J}$ rotate as a vector under this $SU(2)$ symmetry, at the critical 
point each component must have the same scaling dimension.  
As noted in the previous section, the vector $\bm{J}$ and the vector of 
bosonic bilinears $\bm{K}$ have identical symmetry properties under all 
of the microscopic lattice and internal symmetries, and can thus be 
equated at criticality.  So we then ask:  At which of the three
putative critical points in the original spin model does $\bm{K}$ 
transform as an $SU(2)$ vector?

Recall that the transition between the paramagnet and the 
collinearly-ordered spin state is described by two independent 
$U(1)$ models, one for each of the two complex fields $\Phi_{R/L}$.
Since $K_+ = \Phi_R^* \Phi_L$, it is evident that 
$K_z = |\Phi_R|^2 - |\Phi_L|^2$ most definitely does {\it not} have the 
same scaling dimension as $K_x$ and $K_y$ at this transition.  
Moreover, at the transition 
from the paramagnet into the coplanar spin state with long-range 
chiral order, one expects the chirality $\kappa \sim K_z$ correlator to 
decay substantially more slowly than the bond density wave operator 
$B \sim K_+$, which only orders with collinear spin order.
This leaves only the multicritical point in the spin model as a 
candidate fixed point described by QED3 with the full $SU(2)$.

Since the multicritical point is the special point where the paramagnet 
merges with both the coplanar and the collinear spin-ordered states, 
one expects both $K_z$ and $K_\pm$ to have slowly-decaying correlators.
In fact, as the Landau-Ginzburg-Wilson analysis demonstrates, 
the multicritical point 
has an emergent global $O(4)$ symmetry.  Moreover, the three components 
of $\bm{K}$ together with six other bosonic bilinears transform as a 
symmetric and traceless second-rank tensor representation of $O(4)$.
Indeed, the vector $\bm{K}$ transforms as a vector under an $SU(2)$ 
subgroup of the $O(4)$.  We are thus led to the rather bold conjecture:
\emph{The critical theory described by QED3 with two Dirac spinors is 
identical to the $O(4)$ critical point of $\Phi^4$ scalar field theory.}

As detailed in Sec.~\ref{sec:DirectBoson}, at the $O(4)$ critical point 
of scalar field theory, the operators can be arranged into multiplets 
which transform under irreducible representations of the $O(4)$
symmetry group.
For example, the real and imaginary parts of $\Phi_{R/L}$, denoted 
$\chi_i$ with $i=1,\dots,4$ in Eq.~(\ref{O4vector}), transform as a 
vector under $O(4)$, while the ten independent bilinears formed from 
$\chi_i \chi_j$ decompose into an $O(4)$ scalar, 
$K_0 = \sum_{i=1}^4 \chi_i^2$, and the nine components of a symmetric,
traceless matrix that transforms as a second-rank tensor under $O(4)$.  
Moreover, at the $O(4)$ multicritical point of the spin model there are 
precisely two relevant operators that must be tuned to zero: the 
quadratic mass term
\begin{equation}
  \hat{\cal Q}_0 \equiv K_0 = |\Phi_R|^2 + |\Phi_L|^2 ,
\end{equation}
which is an $O(4)$ scalar, and the quartic term
\begin{equation}
  \hat{\cal Q}_2 \equiv \bm{K}^2 - 3K_z^2 ,
\end{equation}
which transforms as spin-2 under $O(4)$ and 
breaks the global $O(4)$ symmetry down to $U(1) \times U(1)$.  
When the coupling of $\hat{\cal Q}_2$ is positive in the Hamiltonian, 
it favors the coplanar spin-ordered state ($\la K_z \ra \ne 0$) over the 
collinear states ($\la K_{x,y} \ra \ne 0$), and vice versa when it is 
negative.

In order to back up our bold conjecture that the critical QED3 theory 
describes this $O(4)$ multicritical point, it is clearly necessary, 
at the very least, to find operators in the QED3 theory that can be 
associated with each of the $O(4)$ multiplets mentioned above.  
That is, one must find operators in the QED3 theory which under the 
microscopic symmetries transform identically to their Landau theory 
counterparts.  Moreover, one should identify the two relevant operators 
in the QED3 theory that transform identically to $\hat{\cal Q}_{0,2}$ 
above.  Completing the latter task requires revisiting the 
issue of stability of the QED3 fixed point.

\subsection{Stability of the QED3 Fixed Point}
\label{subsec:QEDStabl}
To assess the stability of the QED3 fixed point, we need to determine 
which of the quartic interactions are relevant perturbations.  
Equivalently, we need to see which of the quartic interactions have 
scaling dimensions smaller than $D=3$.  In general this is a very 
difficult task.  The best we can do is to examine the trends in the 
$1/N$ expansion.  To leading order in $1/N$ only the $SU(2)$ scalar 
$\hat{Q}_0$ in Eq.~(\ref{Q0}) has a scaling dimension which is reduced 
below $4$, and if we na\"{\i}vely put $N=1$ we find $\Delta_{Q_0} = 0.76 < 3$.
Let's assume that $\hat{Q}_0$ is indeed relevant at the $N = 1$ critical 
QED3 fixed point.  Can we then identify $\hat{Q}_0$ with one of the two
relevant perturbations at the $O(4)$ Landau critical point?  
Since $\hat{Q}_0$ is invariant under all of the symmetries,
it has the same symmetry as the Landau mass term, 
$\hat{\cal Q}_0 = K_0$.  Moreover, if we add $\hat{Q}_0$ to the 
fixed-point QED3 Lagrangian with a positive coupling the 
paramagnetic spin state with $\la J_0 \ra \ne 0$ is favored, whereas a negative
coupling drives magnetic spin order, $\la \bm{J} \ra \ne 0$.
Likewise, a positive mass term $\hat{\cal Q}_0$ when added to the $O(4)$
symmetric fixed point of the Landau action drives one into the 
paramagnet, while a negative mass leads to magnetic order.  
Apparently, it is then entirely consistent to identify the mass term  
$\hat{\cal Q}_0$ in the Landau theory with the QED3 perturbation 
$\hat{Q}_0$.

Similarly, it is possible to identify the second relevant perturbation 
at the $O(4)$ Landau fixed point, $\hat{\cal Q}_2 = \bm{K}^2 -3K_z^2$,
with the quartic perturbation $\hat{Q}_2 = \bm{J}^2 - 3J_z^2$ in QED3.
We have already established that the two vectors $\bm{K}$ and 
$\bm{J}$ transform identically under all of the microscopic symmetries 
(see tables~\ref{tab:phi} and \ref{tab:psi}).  
Moreover, adding the two operators to their respective Lagrangians 
breaks the energetic degeneracy between the coplanar and collinear 
spin-ordered phases.  If this identification is indeed correct, it would imply
that the quartic perturbation $\hat{Q}_2$ is (weakly) relevant when added
to the critical QED3 Lagrangian.  Based on our leading-order $1/N$ 
calculation this is surprising since we found that the leading 
correction increased the scaling dimension of this operator: 
$\Delta_{Q_2} = 4 + 4g/(3\pi^2) + O(g^2)$.  But it is conceivable that
higher-order corrections are negative and dominate when $N=1$.  

At this stage our precise operator correspondence between $O(4)$ scalar 
field theory and QED3 has been limited to operators which are invariant 
under the global $U(1)$ spin symmetries.  In order to complete the task,
then, we must identify operators within QED3 that are 
symmetry-equivalent to the spin fields, $\Phi_{R/L}$, 
as well as the anomalous bilinears, 
$\bm{I} = \Phi_a [\bm{\tau} \tau^y]_{ab} \Phi_b$ and $\bm{I}^\dagger$.
Since these operators carry non-zero spin, they correspond to operators 
in QED3 which add (dual) gauge flux ($\Delta \times a$) in discrete 
units of $2\pi$.  We study the properties of such ``monopole" operators 
in the next subsection.

\subsection{Monopole Operators in QED3}
It is important to note that since we are dealing with a
noncompact gauge theory, we do not have monopoles as dynamical
degrees of freedom in the system.  We can still, however, consider
states with monopoles inserted by hand.
Our treatment is motivated by the work of Borokhov\etal\cite{BKW}, who discussed
such topological disorder operators in three-dimensional QED.

Let us first consider the monopole operators in our dual QED3
formulation that correspond to $\Phi_{R/L}$ and $\Phi_{R/L}^*$.  
Since these fields carry spin $\pm 1$, the corresponding monopole 
operators add $\pm 2\pi$ gauge flux.  
In what follows, we will treat this gauge flux as a classical, 
static background to which the fermions respond.  Moreover, we will 
assume that the flux is spread out over a large area compared to the 
lattice unit cell since these configurations have low energy.  In the
presence of such monopoles, the Hamiltonian density in 
Eq.\ (\ref{contH}) generalizes to
\begin{equation}
  {\mathcal H}_{f,q}^{(0)} = \psi_{a}^\dagger \; 
  [ (p_x - a^q_{x}) \sigma^x +
    (p_y - a^q_{y}) \sigma^y ] \psi_{a} ~,
  \label{contHq}
\end{equation}
where as before $a = R/L$ denotes the fermion flavor and is implicitly
summed.  In Eq.\ (\ref{contHq}), $a^q_{x,y}$ denotes the vector
potential giving rise to $2\pi q$ flux, where $q\in \mathbb{Z}$ is the
monopole ``charge''.
We will define monopole creation and annihilation operators by
specifying their action on the zero-flux state, and
therefore focus on the $q = \pm 1$ sectors.  

In the presence of a charge $q = \pm 1$ monopole, the above Hamiltonian 
admits two zero-energy states, one for each fermion 
flavor.\cite{ZeroModes}  Moreover, one of these states must be
occupied in order to obtain a physical (i.e., gauge-invariant)
$q = \pm 1$ ground state.  
To see this explicitly, first note that since the honeycomb lattice is 
bipartite, sending $d_{\bf x} \to -d_{\bf x}$ on one of the two 
sublattices changes the sign of the fermion hopping part of the lattice
Hamiltonian, Eq.\ (\ref{VortFermHop}).  Hence there is a one-to-one
correspondence between states with energy $E$ and $-E$.  Since the 
fermions are at half-filling, the lowest-energy $q = \pm 1$ 
states will therefore have all 
of the $E<0$ modes and one of the two zero modes occupied.

Thus, we see that our dual QED3 by its own dynamics has two species of
monopole insertions carrying $R/L$ momentum (see below).
Such topological disorder operators are expected to have
nontrivial scaling dimensions in the interacting conformal
field theory.\cite{BKW}
We can also ask how these monopoles transform under the various
microscopic symmetries.  For the lattice symmetries, we can guess
the answer from the transformation properties of the $\psi_{R/L}$ 
fermion fields in Table~\ref{tab:psi} by simply retaining the $\tau$ 
matrix structure acting on the $R/L$ flavor indices.
However, the transformations of the fermion fields have $U(1)$ ambiguity,
while the monopole insertions are gauge-invariant objects and therefore 
have definite transformation properties.  To fix this ambiguity we need
to examine the monopole states more carefully, which is outlined below.

The analysis is performed in a fixed gauge and starts by considering
the zero modes and how they transform under the microscopic symmetries.
We obtain the zero-energy states of Eq.\ (\ref{contHq})
in the presence of a charge $q = \pm 1$ monopole 
following the approach of Jackiw in Ref.\ \onlinecite{ZeroModes}.  
In the Coulomb gauge, the vector potential may be written in terms of 
a scalar function $\alpha_q$ as
$a^q_{i} = \epsilon_{ij} \partial_j \alpha_q$. 
Assuming an azimuthally-symmetric flux distribution centered 
around the origin, we can choose $\alpha_q \sim -q \ln |{\bf x}|$ as 
${\bf x} \to \infty$ to achieve a total gauge flux of $2 \pi q$.
Focusing only on the unit-strength monopoles, we make the 
replacement
\begin{equation}
  \psi_{a}({\bf x}) \rightarrow \phi_{aq}({\bf x}) f_{aq}
  \label{ZeroModes1}
\end{equation}
in Eq.\ (\ref{contHq}),
where for each $a=R/L$ and $q=\pm 1$, $\phi_{aq}({\bf x})$ is the
corresponding two-component, zero-energy wavefunction and 
the operator $f_{aq}$ annihilates this state.
It is straightforward to show\cite{ZeroModes} that with the above 
gauge choice, the only suitable wavefunctions are
\begin{eqnarray}
  \phi_{a,+1} &\sim& \frac{1}{|{\bf x}|} \binom{1}{0} ~,
  \\
  \phi_{a,-1} &\sim& \frac{1}{|{\bf x}|} \binom{0}{1}.
\end{eqnarray}
These states are quasi-localized since their normalization
integrals diverge logarithmically with system size,
but the only information that we actually need is the spinor structure
of these wavefunctions.

The symmetry transformation properties of the zero-mode operators
$f_{aq}$ can be deduced from Eq.\ (\ref{ZeroModes1}) together with the
transformation properties of the continuum fields $\psi_a$ in Table
\ref{tab:psi}.  Under translations for instance, 
$\psi_R \to e^{i {\bf Q}\cdot {\bf \delta r}} \psi_R$, which implies 
that $f_{Rq} \to e^{i {\bf Q}\cdot {\bf \delta r}} f_{Rq}$.  
Since the flux changes sign under particle-hole and fermionic
time reversal, one must additionally transform $q\to -q$ under these
symmetries.  Thus, particle-hole transforms 
$\psi_R \to \sigma^x [\psi_L^\dagger]^t$ and 
$f_{R,q} \to f_{L,-q}^\dagger$, while time reversal sends 
$\psi_R \to i\sigma^y \psi_L$ and $f_{R,q} \to q f_{L,-q}$.
The transformation properties of $f_{R/L, q}$ are summarized in
Table \ref{tab:f}.

\begin{table} 
\caption{\label{tab:f} Transformation properties of the zero-mode
operators $f_{R/L,q}$ in the charge $q = \pm 1$ monopole sectors.
Note that these transformations were obtained by employing the Coulomb gauge.} 
\begin{ruledtabular} 
\begin{tabular}{c | c | c | c | c || c} 
  & $T_{\bf \delta r}$ & $R_{\pi/3}$ & $\tilde {\mathcal R}_x$ & ${\mathcal C}$  & ${\mathcal T}_{\rm ferm}$ \\ 
  \hline 
  $f_{R,q} \to$  
  & $e^{i{\bf Q\cdot \delta r}} f_{R,q}$ 
  & $i e^{-i q \pi /6 } f_{L,q}$  
  & $f_{R,q}$
  & $f_{L,-q}^\dagger$  
  & $q f_{L,-q}$ \\
  \hline 
  $f_{L,q} \to$  
  & $e^{-i{\bf Q\cdot \delta r}} f_{L,q}$
  & $i e^{-i q\pi /6 } f_{R,q}$
  & $f_{L,q}$
  & $f_{R,-q}^\dagger$
  & $-q f_{R,-q}$
  \end{tabular} 
\end{ruledtabular} 
\end{table}

With the zero modes in hand, we are now in position to 
introduce the monopole operators that correspond to $\Phi_{R/L}$ and
$\Phi_{R/L}^*$.
First, we define operators $M^\dagger_{R/L}$ that insert $+2\pi$ flux
and add momentum $\pm{\bf Q}$ by filling the resulting $R/L$ zero mode.  
Their Hermitian conjugates $M_{R/L}$ insert $-2\pi$ flux and add 
momentum $\mp{\bf Q}$ by filling the $L/R$ zero mode.  
In particular, $M^\dagger_{R/L}$ and $M_{R/L}$ act on the zero-flux 
ground state $|0\rangle$ according to
\begin{eqnarray}
  M^\dagger_{R/L} |0\rangle &\equiv& 
  e^{i\alpha_{R/L}}f^\dagger_{R/L,+1}|DS,+1\rangle
  \label{Mdagger}
  \\
  M_{R/L} |0\rangle &\equiv& 
  e^{i\beta_{R/L}}f^\dagger_{L/R,-1}|DS,-1\rangle,
  \label{M}
\end{eqnarray}
where $|DS,\pm 1\rangle$ denotes the filled negative-energy Dirac sea in 
the presence of $\pm 2\pi$ gauge flux, with both zero modes vacant.
The right-hand sides of the above equations are gauge-invariant 
quantum states and can be chosen with arbitrary phases 
$\alpha_{R/L}$ and $\beta_{R/L}$
(indeed, for any orthogonal $|0\ra, |X\ra, |Y\ra$, we can define
$\hat{A} \equiv |X\ra \la 0| + |0\ra \la Y|$ such that
$\hat{A} |0\ra = |X\ra$ and $\hat{A}^\dagger |0\ra = |Y\ra$).
We emphasize that we have defined the monopole creation/annihilation 
operators by constructing the corresponding physical states with
one monopole/antimonopole, and we have the full freedom to specify the 
phases $\alpha_{R/L}$ and $\beta_{R/L}$ to our convenience.

The transformation properties of $M_{R/L}$ can be inferred from
Eqs.\ (\ref{Mdagger}) and (\ref{M}), provided we know how the states
$|0\rangle$ and $|DS, \pm 1\rangle$ transform.  While $|0\rangle$
transforms simply,
deducing the symmetry properties of the negative-energy Dirac sea in 
the presence of $\pm 2\pi$ flux is challenging since this would 
require the knowledge of all wavefunctions and not only the zero-energy
modes.  Nevertheless, even without such knowledge, by employing
relations among the symmetries we can infer the needed transformation 
properties of $|DS,q\rangle$ so as to be able to fix the transformation
properties of $M_{R/L}$ almost completely by appropriate choices
of $\alpha_{R/L}$ and $\beta_{R/L}$.
As this is somewhat involved, we relegate the details to 
Appendix \ref{app:DS},
and simply recall the results here.  We find that the filled 
negative-energy Dirac sea in the presence of a charge 
$q = \pm 1$ monopole transforms as follows:
\begin{eqnarray}
  T_{\bf \delta r}|DS,q\rangle &=& |DS,q\rangle
  \nonumber \\
  R_{\pi/3}|DS,q\rangle &=& \zeta_{\pi/3} e^{-i q 2\pi/3}|DS,q\rangle
  \nonumber \\
  \tilde {\mathcal R}_x|DS,q\rangle &=& \zeta_x|DS,q\rangle
  \label{DStrans}
  \\
  {\mathcal C} |DS,q\rangle &=&
  f^\dagger_{R,-q}f^\dagger_{L,-q}|DS,-q\rangle
  \nonumber \\
  {\mathcal T}_{\rm ferm}|DS,q\rangle &=& |DS,-q\rangle.
  \nonumber
\end{eqnarray}
Here $\zeta_{\pi/3}, \zeta_x= \pm 1$ are possible overall signs 
which have not been determined in our analysis.

With the help of Table~\ref{tab:f}, we now know the transformation 
properties of the gauge-invariant states in Eqs.\ (\ref{Mdagger}) and 
(\ref{M}), and it is easy to check that we can choose the phases 
$\alpha_{R/L}$ and $\beta_{R/L}$ such that the monopole operators 
transform as follows,
\begin{eqnarray}
  T_{\delta {\bf r}} &:& M_{R/L} \rightarrow e^{\pm i {\bf Q}\cdot 
  {\delta {\bf r}}}M_{R/L} 
  \nonumber \\
  R_{\pi/3} &:& M_{R/L} \rightarrow M_{L/R}
  \nonumber \\
  \tilde {\mathcal R}_x &:& M_{R/L} \rightarrow M_{R/L}
  \label{Mtrans} \\
  {\mathcal C} &:& M_{R/L} \rightarrow M_{L/R}^\dagger
  \nonumber \\
  {\mathcal T}_{\rm ferm} &:& M_{R/L} \rightarrow s M_{R/L}^\dagger.
  \nonumber
\end{eqnarray}
Here the sign $s=\pm 1$ is in principle fixed, but can not be
determined from our analysis.
Comparing with Table~\ref{tab:phi} we see that we have constructed
monopole insertion operators $M_{R/L}$ in the dual QED3 that 
transform identically to $\Phi_{R/L}$ in the spin system under the 
lattice symmetries and particle-hole.  Additionally, $M_{R/L}$
transform under ${\cal T}_{\rm ferm}$ in the same way that
$\Phi_{R/L}$ transform under the physical time reversal, up to a
possible overall minus sign.
We are thus content with the agreement obtained here between the 
discrete symmetries in the two systems (a more precise analysis
would of course be welcome).

We turn next to double-strength, charge $q = \pm 2$ monopoles.  
Although the operators $M_{R/L}$ defined above are insufficient 
to create all such monopoles (see below), we can nevertheless 
deduce from $M_{R/L}$ the transformation properties of the 
\emph{most relevant} double-strength monopole operators using the 
following argument.  For example, one can use $M_{R/L}^\dagger$ to 
create two isolated $q = +1$ monopoles separated by a large distance.  
If we fuse these monopoles together, the quantum numbers of the 
resulting double-strength monopole will be the sum of those for 
the two unit-strength monopoles since the quantum numbers are conserved.  
Such double-strength monopole operators must therefore transform 
identically to the ``bilinears'' $M_R^2$, $M_L^2$, and $M_R M_L$,
where we treat $M_{R/L}$ as bosonic fields.
These bilinears can be grouped into an $SU(2)$ vector,
\begin{equation}
  \bm{D} = M_a[\bm{\tau} \tau^y]_{ab} M_b,
\end{equation}
which transforms identically to the $SU(2)$ vector $\bm{I}$ defined in 
Eq.\ (\ref{Ivec}).  
Thus, we have established the correspondence between the anomalous 
spin-field bilinears and double-strength monopole operators.  

As remarked above, the preceding argument pertains only to the most
relevant double-strength monopoles.
In the presence of $4\pi$ flux there are four zero modes
(two for each flavor), and two of these need to be occupied to obtain 
a gauge-invariant state.  Hence, one has to consider \emph{six} 
double-strength monopole operators since there are six ways of 
filling the zero modes.  
However, three of these states are not locally gauge-neutral and 
have contributions to the gauge charge density with dipolar angular
dependence.  Such states have higher energy due to associated 
electric fields than the remaining three states which have 
rotationally-invariant locally neutral charge distributions.  
Operators that create such dipolar states are expected to have 
higher scaling-dimensions and are neglected, leaving only the three 
double-strength monopole operators considered above.  

To summarize our discussion, the monopole insertions $M_{R/L}$
in QED3 correspond to the $\Phi_{R/L}$ fields in the spin model.
We predict that the scaling dimension of such monopole operators 
in QED3 with $N=1$ and full $SU(2)$ flavor symmetry is given by 
the scaling dimension of the ordering field $\Phi$ at the $O(4)$ 
critical point of $\Phi^4$ scalar field theory.
Furthermore, we identified double-strength monopoles 
$\bm{D}, \bm{D}^\dagger$ in QED3 that correspond to the 
spin-field bilinears $\bm{I}, \bm{I}^\dagger$.
Recall also that the fermionic bilinears $\bm{J}$ correspond to 
the spin bilinears $\bm{K}$.  At the $O(4)$ fixed point,
$\bm{K}$, $\bm{I}$, and $\bm{I}^\dagger$ all have the same
scaling dimension---these are the quadratic anisotropy
fields of the $\Phi^4$ theory.  Thus, we are led to a rather unusual
prediction: \emph{the fermionic bilinears and the double-strength 
monopole insertions have the same scaling dimension in our 
QED3 theory}.  We emphasize that this is striking from
the perspective of the large-$N$ limit, where the monopole
operators have large scaling dimension, gauge fluctuations being
strongly suppressed, while the fermionic bilinears have scaling
dimension $2 - {\cal O}(1/N)$.  It would be quite remarkable if the scaling dimensions
of these operators which have no obvious relation 
indeed merge in the physically-relevant $N = 1$ case.

\subsection{Dual Picture for the Honeycomb QED3}
\label{subsec:QEDDual}
We conclude this section with a separate argument for the applicability
of the critical QED3 theory to the description of the continuous phase
transitions in the original spin model.  We intend to further address
the question of whether it is appropriate to drop the higher-derivative
Chern-Simons-like term in the fermionized-vortex Lagrangian Eq.~(\ref{Lfinal}) as
irrelevant at criticality and at the same time prohibit the mass term 
$M\barpsi\psi$.
The setting below, which in many respects is a reversal of the preceding
line of attack (but \emph{not} a circular argument), gives us good
control on the effects of the Chern-Simons gauge field in a situation 
where the particles in the theory already have strong interactions 
mediated by a photon field.  
Our difficulties in Sec.~\ref{subsec:FermCrit}, which we distilled to 
the question regarding the realization of the physical spin time-reversal 
symmetry in the continuum dual action, stem from the inability to treat 
the statistical interactions accurately in this context.
Specifically, the local symmetries of the formal Lagrangian 
Eq.~(\ref{Lfinal}) allow a generic mass term $M\barpsi\psi$ which
is of course a very relevant perturbation in the Dirac field theory.
On the other hand, all such ${\cal T}_{\rm ferm}$-violating terms
have to conspire to recover the original spin time-reversal invariance
which appears lost in the continuum theory.

We therefore look for a similar situation where the statistical 
interactions can be treated accurately.  In the present context,
it is natural to consider the honeycomb lattice QED3 problem 
\emph{per se}, and seek a ``dual'' description of this 
fermionic theory in a manner 
familiar from the treatment of ``vortices'' in fractional 
quantum Hall systems.\cite{FQHvort1,FQHvort2,FQHvort3}
This is achieved by viewing the lattice fermions as bosons 
carrying fictitious $2\pi$ flux and performing a duality 
transformation on these Chern-Simons bosons.  The Chern-Simons 
bosons are governed by a 
Hamiltonian similar to Eq.~(\ref{dualH2}), but with additional 
statistical interactions mediated by the corresponding Chern-Simons 
gauge field.  The duality transformation essentially reverses the steps 
in Sec.~\ref{sec:BosVort}, but in the presence of the Chern-Simons gauge field 
coupled to the currents conjugate to the $\theta$ fields.  
In the absence of the Chern-Simons gauge field, we would of course recover the 
original XY spin model on the frustrated triangular lattice.  
For example, in terms of the integer-valued currents $j$ conjugate to 
the $\phi$ fields, we would obtain a Euclidean action of the form
\begin{equation}
S_{XY}[j] = \sum \frac{j^2}{2\beta} 
- i \sum j_{\bf rr'} {\cal A}^0_{\bf rr'} ~,
\end{equation}
where the first term denotes schematically generic short-range repulsion 
between the currents and the second term encodes the frustration for the 
$\phi$-boson hopping.  Including the Chern-Simons gauge field, 
the corresponding 
integration can be performed essentially exactly and leads to
a new contribution of the form
\begin{equation}
\delta S[j] = i \kappa \sum j \cdot (\nabla \times j)
\end{equation}
with some nonuniversal numerical coefficient $\kappa$.
Thus, the dual of the honeycomb lattice QED3 model is the stacked 
triangular lattice antiferromagnetic XY spin model with additional 
current-current interactions that break spin time-reversal and
$x$-reflection symmetry.

The advantage here compared with our fermionized vortex system
of Sec.~\ref{sec:FermVort} is that the new interactions $\delta S[j]$
are \emph{local} in terms of the spin fields of the new spin model.  
We also observe that the new model still respects the translation 
($T_{\delta \bf r}$), rotation ($R_{\pi/3}$), 
\emph{modified} reflection ($\tilde{\cal R}_{x}$), and particle-hole 
(${\mathcal C}$) symmetries defined in Sec.~\ref{subsec:DirectLGW}.
Furthermore, the expected charge-ordered and integer quantum Hall
fermionic states of the honeycomb QED3 model are recovered as, 
correspondingly, the spin-ordered and paramagnetic phases in the 
dual description.
Performing the continuum analysis of the new spin model,
we conclude that the above symmetries preclude all relevant
terms other than the ones already exhibited in the action
Eq.~(\ref{Seff}).  In particular, the additional terms that violate 
spin time-reversal symmetry necessarily have more derivatives
and cannot generate new relevant terms.  We therefore conclude
that these terms are irrelevant in the $\Phi^4$ theories that
govern criticalities in the spin model, and that the spin
time-reversal symmetry emerges at the corresponding fixed
points.  Thus, it appears that we have identified the continuum 
$N=1$ massless QED3 with the criticality in our original classical 
spin system discussed in detail in Sec.~\ref{sec:DirectBoson}.
A possible flaw in this identification is that the new interactions 
$\delta S[j]$ are numerically very large, and therefore put the 
spin system corresponding to the lattice QED3 into some very different 
regime of the phase diagram.
We cannot rule this out with the present approach, but our predictions
for the QED3 criticality can be in principle tested in direct
lattice simulations.

\section{Conclusion}
\label{sec:Conclusion}

In this paper we have introduced a new dual vortex approach to
easy-plane frustrated quantum spin systems which is well-suited 
for the study of novel critical points and critical phases.  
Specifically, we demonstrated how one can formulate a low-energy dual
description of frustrated systems by fermionizing the vortices, 
which are at finite-density due to frustration, 
via Chern-Simons flux attachment.  A detailed 
analysis was carried out for the easy-plane integer-spin 
triangular antiferromagnet, guided by the known results from a 
direct Landau analysis of the spin model.  The fermionized-vortex
approach led naturally in this case to a low-energy Dirac theory 
for two species of
fermions coupled to a Chern-Simons field and a $U(1)$ gauge field
that mediates logarithmic vortex interactions.  We demonstrated how
the ordered phases of the spin model are captured in the continuum 
dual theory via spontaneous fermion mass generation, and conjectured 
that the critical QED3 theory obtained by essentially ignoring the 
Chern-Simons field describes the $O(4)$ multicritical point of 
$\Phi^4$ theory.  
This rather bold conjecture led us to some surprising yet 
testable predictions for the scaling dimensions
of various operators in QED3.  Future work, such as lattice
simulations of QED3\cite{Kogut} and a higher-order renormalization group
analysis assessing the stability of the QED3 fixed point, 
may provide further insight into the validity of this correspondence.

To put our result in perspective, we have established duality between 
the two interacting conformal field theories by considering microscopic
models, transforming to appropriate topological disorder variables,
and studying detailed correspondence of the phases and transitions in 
the two systems.
Our approach may also be useful in other contexts where such 
dualities have been conjectured.\cite{Aharony, BKW2}

Another 
physically interesting setting in which this dual approach
can be applied is the easy-plane spin-1/2 triangular 
antiferromagnet.  In this case a
direct Landau approach is \emph{not} accessible, while
the techniques introduced here permit an analysis in the dual
language.  Such an analysis can provide insight into critical 
spin-liquid phases, and will be carried out in a forthcoming paper. 
This may be useful for understanding recently observed spin liquids 
in the quasi-2D spin-1/2 triangular antiferromagnets\cite{Coldea,Kanoda} 
$\text{Cs}_2\text{CuCl}_4$ 
and $\kappa\text{-(ET)}_2\text{Cu}_2(\text{CN})_3$.
The easy-plane kagome antiferromagnet is another example of a 
frustrated spin system that might be fruitfully explored within a 
dual fermionic formulation.  
More generally, we expect that fermionization can provide an 
effective tool for describing criticality of bosons at finite density 
in situations where the bosons are strongly-interacting,
\cite{Feigelman, DHKim}
leading to a diminished role of exchange statistics.

\begin{acknowledgments} 
This work was supported by the Department of Defense NDSEG program 
(M.\ H.) and the National Science Foundation through a Graduate 
Research Fellowship (J.\ A.) and grants PHY-9907949 (O.~I.~M.\ and  
M.\ P.\ A.\ F.) and DMR-0210790 (M.\ P.\ A.\ F.).   
 
\end{acknowledgments}

\appendix
\section{Application of fermionized vortices to bosons at $\nu=1$}
\label{app:nu1}
In this Appendix, we provide some details of the fermionized vortex
approach to bosons at filling factor $\nu=1$ discussed in 
the Introduction.
This proceeds as for the easy-plane spins, which in fact we viewed 
as bosons in a magnetic field. 
To be closer to the continuum, we now imagine bosons say on a 
square lattice at low average density $\bar\rho \ll 1$ per site.
The dual theory in the present $\nu=1$ case has bosonic vortices 
at filling factor $\nu_{\rm dual} = 1/\nu = 1$.  Vortices are
at finite density because of the original magnetic field, while
the dual magnetic field arises because the original bosons are
at finite density.
However, unlike the original bosons, vortices interact via a 2D 
electromagnetic interaction, and fermionization of vortices
is more controlled.  Upon Chern-Simons flux attachment and 
taking a flux-smeared mean-field state in which the statistical flux
cancels the dual magnetic field, 
we obtain fermionic vortices in zero average field
coupled to $U(1)$ gauge field and Chern-Simons gauge field.
This theory is very similar to Eq.~(\ref{Lintro}), except that 
vortices are nonrelativistic at finite density $\bar\rho$, 
which is the same as the density of the original bosons.  
The fermionic vortex Lagrangian reads
\begin{eqnarray} 
{\mathcal L}_{\nu=1} &=& 
\psi^\dagger \left(\partial_\tau 
                   -\frac{[{\bm \nabla} - i ({\bm a} + {\bm A})]^2}
                         {2 m_{\rm vort}}
             \right) \psi 
\nonumber \\
&-& i (a_0 + A_0) (\psi^\dagger \psi - \bar\rho)
\nonumber \\ 
&-& \frac{i}{4\pi} A \cdot \nabla \times A
 - \frac{i}{2\pi} A^{\rm ext} \cdot \nabla \times a ~.
\label{Lnu1}
\end{eqnarray}
Here $m_{\rm vort}$ is a vortex mass which we treat as a
phenomenological parameter.  To allow computation of physical
response properties, we have included an external probing field 
$A^{\rm ext}$ minimally coupled to the original boson 
three-current $\delta j = \frac{\nabla \times a}{2\pi}$.
We also need to add the following bare action for the dual
gauge field $a$:
\begin{eqnarray*}
S_{\rm bare}[a] &=& \frac{m_0}{2 (2\pi)^2} \int d\tau\, d{\bf r}\, 
({\bm \nabla} a_0 - \partial_\tau {\bm a})^2
\nonumber \\
&+& \frac{1}{2 (2\pi)^2} \int d\tau\, d{\bf r}\, d{\bf r'}\, 
\,v_{\bf rr'}\, (\bm{\nabla \wedge a})_{\bf r} 
                                 (\bm{\nabla \wedge a})_{\bf r'} ~.
\end{eqnarray*}
Here $m_0$ is roughly the original boson mass and $v_{\bf rr'}$
encodes the boson density-density interaction.  The limit $m_0 \to 0$
corresponds to restricting the original bosons to the lowest 
Landau level.
With the above ingredients and using an RPA approximation for integrating
out the fermions, we can, e.g., obtain the physical response properties
of the compressible state in agreement with Ref.~\onlinecite{Read}.

From the above action, we see that the statistical gauge field $A$ is
`screened' by the dual gauge field $a$.  More formally, changing to
$\tilde{a} = a + A$ and integrating over the field $A$, we obtain
\begin{eqnarray} 
{\mathcal L}_{\nu=1} &=& 
\psi^\dagger 
  \left(\partial_\tau 
        -\frac{[{\bm \nabla} - i \tilde{\bm a}]^2}{2 m_{\rm vort}}
  \right) \psi 
- i \tilde{a}_0 (\psi^\dagger\psi - \bar\rho) + \dots ~,
\label{Lstrongcoupl}
\end{eqnarray}
which is the equivalent of Eq.~(\ref{Lfinalintro}) for nonrelativistic
vortices.  For simplicity, we do not write the quadratic terms for the
gauge field $\tilde{a}$ that arise from $S_{\rm bare}[a]$.
For finite $m_0$ and short-range $v_{\bf rr'}$ these have the
general structure shown in Eq.~(\ref{Lfinalintro}).

We now specialize to the lowest Landau level limit $m_0 \to 0$.
The final action then describes fermions coupled to the gauge field 
$\tilde{a}$ with \emph{no} Chern-Simons term, which is similar to
the result of Ref.~\onlinecite{Read}.

Observe that in the present formulation the fermions are neutral by 
construction and do not couple directly to the probing field 
$A^{\rm ext}$.  However, as the following calculation shows, 
fermions carry \emph{electrical dipole moments} oriented perpendicular 
to their momentum ${\bm k}$ and of strength proportional to $|{\bm k}|$.
Indeed, from the action Eq.~(\ref{Lstrongcoupl}) we obtain the following
equations of motion,
\begin{eqnarray}
\psi^\dagger \psi &=& \bar\rho ~,\\
\tilde{\bm a}\, \psi^\dagger \psi 
&=& -\frac{i}{2} [\psi^\dagger {\bm\nabla}\psi - 
               ({\bm\nabla} \psi^\dagger) \psi] ~.
\end{eqnarray}
(Incidentally, the first equation simply means that in the $m_0 \to 0$
limit the field $a_0$ is completely soft and its fluctuations fix
the vortex density; clearly, the vortex fermionization procedure
is the simplest here.)
Now, the original boson density is given by
\begin{equation}
 \delta\rho_{\rm bos} = 
    \frac{{\bm\nabla} \wedge \tilde{\bm a}}{2\pi}
  - \frac{{\bm\nabla} \wedge {\bm A}}{2\pi} = 
    \frac{{\bm\nabla} \wedge \tilde{\bm a}}{2\pi}
  + (\psi^\dagger \psi - \bar\rho) ~,
\end{equation}
and using the equations of motion we find
\begin{equation}
\delta\rho_{\rm bos} ({\bm q}) 
= \frac{1}{2\pi\bar\rho} \int \frac{d^2 k}{(2\pi)^2} \;
 i {\bm q} \wedge {\bm k} \; \psi^\dagger_{{\bm k} - {\bm q}/2}
                             \psi_{{\bm k} + {\bm q}/2} ~.
\end{equation}
Writing $\delta\rho_{\rm bos}({\bm q}) = -i {\bm q} \cdot {\bm P}({\bm q})$,
we interpret this as a charge density created by a medium with bound 
charges and polarization
\begin{equation}
{\bm P}({\bm q}) = \int \frac{d^2 k}{(2\pi)^2} \; (-1)
 \frac{\wedge {\bm k}}{2\pi\bar\rho} \; 
 \psi^\dagger_{{\bm k} - {\bm q}/2} \psi_{{\bm k} + {\bm q}/2} ~.
\end{equation}
As claimed, fermions carry dipole moment $\sim \wedge {\bm k}$.
Thus, the fermionized-vortex formulation reproduces the dipole
picture of the $\nu=1$ state of bosons.

\section{Transformation properties of the negative-energy Dirac sea} 
\label{app:DS}
In this Appendix we obtain the transformation properties given in 
Eq.~(\ref{DStrans}) of the filled negative-energy Dirac sea in 
the presence of a charge $q = \pm 1$ monopole, $|DS, q \rangle$.
We first note that the quoted action of ${\cal T}_{\rm ferm}$ 
and ${\cal C}$ represents a convention for the phases of 
$|DS, q \rangle$.
Time reversal sends the negative-energy Dirac sea of a monopole
into the negative-energy Dirac see of an antimonopole, while
particle-hole additionionally fills the two zero-energy modes of the 
antimonopole.  One can thus use the specified transformations
of $|DS, +1\ra$ to define the phases of $|DS, \pm 1\ra$, while
the transformations of $|DS, -1\ra$ follow from 
${\cal T}_{\rm ferm}^2 = 1$ and ${\cal C}^2 = 1$.

We can also give a more explicit construction by considering
some general properties of the single-particle wavefunctions in the
presence of a vector potential $a^q_{\bf x x'}$ that gives rise to
$2\pi q$ flux.  The fermion hopping Hamiltonian is
\begin{equation}
  H_{t,q} = -t_v \sum_{\langle {\bf x} {\bf x'} \rangle}
  [d^\dagger_{{\bf x}} d_{{\bf x'}} e^{-i a^q_{\bf x x'}} + \text{h.c.}].
\end{equation}
This can be formally diagonalized by expanding $d_{{\bf x}}$ in
the energy basis:
\begin{equation}
  d_{{\bf x},\alpha} = \sum_E \phi^E_{q,\alpha}({\bf x}) c_{Eq},
  \label{Ebasis}
\end{equation} 
where $\alpha=1,2$ is the sublattice index, $\phi^E_q({\bf x})$ is a 
two-component wavefunction, and $c_{Eq}$ annihilates the state with
energy $E$ in the presence of $2\pi q$ flux.  
The above sum is over all energy eigenstates.  

With our gauge choice, we have $a^q_{\bf x x'} = -a^{-q}_{\bf x x'}$.
Replacing $d_{{\bf x},\alpha} \rightarrow 
\sigma^z_{\alpha\beta} d^\dagger_{{\bf x},\beta}$ therefore sends
$H_{t,q} \rightarrow H_{t,-q}$.  This implies a relation between
charge $q$ and $-q$ wavefunctions,
\begin{equation}
  \phi^E_q({\bf x}) = \phi^{E*}_{-q}({\bf x}).
  \label{qrelation}
\end{equation}
Moreover, the transformation $d_{{\bf x},\alpha} \rightarrow
\sigma^z_{\alpha\beta} d_{{\bf x},\beta}$ sends $H_{t,q} \rightarrow
-H_{t,q}$,
which implies the following relation between wavefunctions with energy
$E$ and $-E$,
\begin{equation}
  \phi_q^{-E}({\bf x}) = \sigma^z \phi_q^E({\bf x}).
  \label{Erelation}
\end{equation}

These properties allow us to determine how the operators $c_{Eq}$, and
in turn $|DS,q\rangle$, transform under the internal symmetries 
${\cal T}_{\rm ferm}$ and ${\mathcal C}$.  
Consider fermionic time reversal first.  According to Eq.~(\ref{Tferm}), 
${\mathcal T}_{\rm ferm}$ sends $d_{\bf x}\to d_{\bf x}$ and $q\to -q$ 
since the flux changes sign.  
We find using Eqs.\ (\ref{Ebasis}) and (\ref{qrelation}) that 
${\mathcal T}_{\rm ferm}$ transforms $c_{E,q}\to c_{E,-q}$.  
According to Table \ref{tab:d}, particle-hole sends 
$d_{{\bf x}\alpha} \rightarrow \sigma^z_{\alpha\beta}
d^\dagger_{{\bf x}\beta}$ and also changes the sign of the flux.
Using Eqs.\ (\ref{Ebasis}) through (\ref{Erelation}), we then find
that $c_{E,q} \rightarrow c^\dagger_{-E,-q}$ under $\mathcal{C}$.

\emph{Fermionic time reversal}.  Define $|vac,q\rangle$ to be the 
state with $2\pi q$ flux and all fermion modes unoccupied.
Since the flux changes sign under time reversal, 
one can choose the phases of the fermion vacua such that 
${\mathcal T}_{\rm ferm}|vac, q\rangle = |vac, -q\rangle$. 
We write the filled negative-energy sea as
\begin{equation}
  |DS,q\rangle = e^{iq\gamma} \prod_{E<0} c_{Eq}^\dagger|vac,q\rangle,
  \label{DS}
\end{equation}
where $\gamma\in \mathbb{R}$ can be selected arbitrarily. 
Putting together the above results, we find that 
\begin{equation}
  {\mathcal T}_{\rm ferm}|DS,q\rangle = |DS,-q\rangle.
\end{equation}

\emph{Particle-hole}.  
Under particle-hole, we expect physically that the vacuum with 
$2\pi q$ flux should transform to a state with $-2\pi q$ flux and 
all energy states filled.  Using $\mathcal{C}^2 = 1$, 
one can write without loss of generality
\begin{equation}
  {\mathcal C}|vac, q\rangle = e^{i q\delta} 
  \prod_E c_{E, -q}^\dagger |vac, -q\rangle
\end{equation}
by appropriately ordering the creation operators on the right side.
Note that $\delta$ is \emph{not} arbitrary since the phases of
$|vac,q\rangle$ were already fixed.  We now choose the phase 
$\gamma$ so that the negative-energy sea transforms under
particle-hole as follows,
\begin{equation}
  {\mathcal C}|DS, q\rangle =
   f^\dagger_{R, -q} f^\dagger_{L, -q} |DS, -q\rangle .
\end{equation}

We are unable to determine how the operators $c_{Eq}$ transform under 
the lattice symmetries without knowing the explicit forms of the 
wavefunctions.  Nevertheless, under reasonable assumptions we can use 
various identities to determine how the negative-energy Dirac sea 
transforms under $T_{\bf \delta r}$, and (up to overall signs) 
$\tilde{\mathcal R}_x$ and $R_{\pi/3}$.

\emph{Translations}.  To deduce how the Dirac sea transforms under 
translations, we assume that $|DS, q\rangle$ is an eigenstate of
the translation and rotation operators,
\begin{eqnarray}
  T_{\bf \delta r} |DS, q\rangle &=& 
    e^{i{\bf p}_q \cdot {\bf \delta r}} |DS, q\rangle 
  \label{Trans}
  \\
  R_{\pi/3} |DS, q\rangle &=& e^{i m_q \pi/3} |DS, q\rangle.
  \label{Rot}
\end{eqnarray}
Here, ${\bf p}_q$ and $m_q$ are the total momentum and angular
momentum, respectively, of the Dirac sea.  Define triangular lattice
vectors ${\bf \delta r}_1 = \hat {\bf x}$ and 
${\bf \delta r}_2 = -\frac{1}{2} \hat {\bf x} 
                    +\frac{\sqrt{3}}{2} \hat {\bf y}$.  
The following operator relations are expected to hold,
\begin{eqnarray}
  T_{{\bf \delta r}_1} R_{\pi/3} &=& R_{\pi/3} T_{{\bf \delta r}_2}^{-1}
  \\
  T_{{\bf \delta r}_2} R_{\pi/3} &=& R_{\pi/3} T_{{\bf \delta r}_1}
  T_{{\bf \delta r}_2}
\end{eqnarray}
since the left and right hand sides transform the lattice
identically.  Requiring these relations to hold on the negative-energy 
Dirac sea, we find that $|DS, q\rangle$ must carry no overall momentum,
i.e.,
\begin{equation}
  T_{\bf \delta r} |DS, q\rangle = |DS, q\rangle.
\end{equation}

\emph{Modified reflections}.  The modified reflection 
$\tilde{\mathcal R}_x$ is a simple coordinate reflection 
composed with time reversal and does not change the direction 
of the flux.  We expect the negative-energy Dirac sea to be an 
eigenstate of this operator.
Furthermore, commutation with particle-hole and time reversal 
can be used to show that the $q = \pm 1$ states have the same 
eigenvalue $\zeta_x$, which can either be $+1$ or $-1$:
\begin{equation}
  \tilde {\mathcal R}_x |DS, q\rangle = \zeta_{x} |DS, q\rangle.
\end{equation}
To determine the sign $\zeta_x$, one would need to know the 
wavefunctions $\phi_q^E$ in Eq.\ (\ref{Ebasis}) to deduce how the 
operators $c_{Eq}$ transform under $\tilde{\mathcal R}_x$.

\emph{Rotations}.  To obtain the angular momentum $m_{q}$ 
defined in Eq.\ (\ref{Rot}), we first note that commutation with
time reversal immediately implies that $m_{-q} = -m_q$.  
We restrict $m_q$ further using the observation that the composite 
operation 
\begin{equation}
  {\mathcal O} \equiv R^{-1}_{\pi/3} {\mathcal C} R_{\pi/3} {\mathcal C}
\end{equation}
sends $d_{{\bf x},\alpha} \rightarrow -d_{{\bf x},\alpha}$ and leaves 
the flux unchanged.  This can be readily verified from Table \ref{tab:d}, 
recalling that $R_{\pi/3}$ interchanges the two sublattices.
Equation~(\ref{Ebasis}) then implies that ${\mathcal O}$ transforms 
$c_{Eq} \to -c_{Eq}$.  The filled Dirac sea is therefore an eigenvector 
of this operation with eigenvalue $(-1)^{N^<_f}$, where 
$N^<_{f}$ is the total number of fermions residing in the filled 
negative-energy sea.  Assuming that gauge-invariant states have even
total fermion number and noting that such states require one of the two 
zero-modes to be occupied, we conclude that in this case 
$N^<_{f}$ is odd.  We thus have 
${\mathcal O} |DS, q\rangle = -|DS, q\rangle$.  This implies that
\begin{equation}
  R_{\pi/3} |DS, q\rangle = \zeta_{\pi/3} e^{-i q 2\pi/3} |DS, q\rangle,
\end{equation}
where $\zeta_{\pi/3} = \pm 1$ is an overall sign we are unable to
determine without knowing the explicit wavefunctions $\phi^E_q$.


\begin{thebibliography}{50}
\expandafter\ifx\csname natexlab\endcsname\relax\def\natexlab#1{#1}\fi
\expandafter\ifx\csname bibnamefont\endcsname\relax
  \def\bibnamefont#1{#1}\fi
\expandafter\ifx\csname bibfnamefont\endcsname\relax
  \def\bibfnamefont#1{#1}\fi
\expandafter\ifx\csname citenamefont\endcsname\relax
  \def\citenamefont#1{#1}\fi
\expandafter\ifx\csname url\endcsname\relax
  \def\url#1{\texttt{#1}}\fi
\expandafter\ifx\csname urlprefix\endcsname\relax\def\urlprefix{URL }\fi
\providecommand{\bibinfo}[2]{#2}
\providecommand{\eprint}[2][]{\url{#2}}

\bibitem[{\citenamefont{Coldea et~al.}(2001)\citenamefont{Coldea, Tennant,
  Tsvelik, and Tylczynski}}]{Coldea}
\bibinfo{author}{\bibfnamefont{R.}~\bibnamefont{Coldea}},
  \bibinfo{author}{\bibfnamefont{D.~A.} \bibnamefont{Tennant}},
  \bibinfo{author}{\bibfnamefont{A.~M.} \bibnamefont{Tsvelik}},
  \bibnamefont{and}
  \bibinfo{author}{\bibfnamefont{Z.}~\bibnamefont{Tylczynski}},
  \bibinfo{journal}{Phys.\ Rev.\ Lett.} \textbf{\bibinfo{volume}{86}},
  \bibinfo{pages}{1335} (\bibinfo{year}{2001}).

\bibitem[{\citenamefont{Coldea et~al.}(2003)\citenamefont{Coldea, Tennant, and
  Tylczynski}}]{Coldea2}
\bibinfo{author}{\bibfnamefont{R.}~\bibnamefont{Coldea}},
  \bibinfo{author}{\bibfnamefont{D.~A.} \bibnamefont{Tennant}},
  \bibnamefont{and}
  \bibinfo{author}{\bibfnamefont{Z.}~\bibnamefont{Tylczynski}},
  \bibinfo{journal}{Phys.\ Rev.\ B} \textbf{\bibinfo{volume}{68}},
  \bibinfo{pages}{134424} (\bibinfo{year}{2003}).

\bibitem[{\citenamefont{Shimizu et~al.}(2003)\citenamefont{Shimizu, Miyagawa,
  Kanoda, Maesato, and Saito}}]{Kanoda}
\bibinfo{author}{\bibfnamefont{Y.}~\bibnamefont{Shimizu}},
  \bibinfo{author}{\bibfnamefont{K.}~\bibnamefont{Miyagawa}},
  \bibinfo{author}{\bibfnamefont{K.}~\bibnamefont{Kanoda}},
  \bibinfo{author}{\bibfnamefont{M.}~\bibnamefont{Maesato}}, \bibnamefont{and}
  \bibinfo{author}{\bibfnamefont{G.}~\bibnamefont{Saito}},
  \bibinfo{journal}{Phys.\ Rev.\ Lett.} \textbf{\bibinfo{volume}{91}},
  \bibinfo{pages}{107001} (\bibinfo{year}{2003}).

\bibitem[{\citenamefont{Chung et~al.}(2001)\citenamefont{Chung, Marston, and
  McKenzie}}]{Chung1}
\bibinfo{author}{\bibfnamefont{C.~H.} \bibnamefont{Chung}},
  \bibinfo{author}{\bibfnamefont{J.~B.} \bibnamefont{Marston}},
  \bibnamefont{and} \bibinfo{author}{\bibfnamefont{R.~H.}
  \bibnamefont{McKenzie}}, \bibinfo{journal}{J.\ Phys.: Condens.\ Matter}
  \textbf{\bibinfo{volume}{13}}, \bibinfo{pages}{5159} (\bibinfo{year}{2001}).

\bibitem[{\citenamefont{Chung et~al.}(2003)\citenamefont{Chung, Voelker, and
  Kim}}]{Chung2}
\bibinfo{author}{\bibfnamefont{C.~H.} \bibnamefont{Chung}},
  \bibinfo{author}{\bibfnamefont{K.}~\bibnamefont{Voelker}}, \bibnamefont{and}
  \bibinfo{author}{\bibfnamefont{Y.~B.} \bibnamefont{Kim}},
  \bibinfo{journal}{Phys.\ Rev.\ B} \textbf{\bibinfo{volume}{68}},
  \bibinfo{pages}{094412} (\bibinfo{year}{2003}).

\bibitem[{\citenamefont{Zhou and Wen}()}]{Wen}
\bibinfo{author}{\bibfnamefont{Y.}~\bibnamefont{Zhou}} \bibnamefont{and}
  \bibinfo{author}{\bibfnamefont{X.~G.} \bibnamefont{Wen}},
  \bibinfo{note}{cond-mat/0210662 (unpublished)}.

\bibitem[{\citenamefont{Isakov et~al.}()\citenamefont{Isakov, Senthil, and
  Kim}}]{Isakov}
\bibinfo{author}{\bibfnamefont{S.~V.} \bibnamefont{Isakov}},
  \bibinfo{author}{\bibfnamefont{T.}~\bibnamefont{Senthil}}, \bibnamefont{and}
  \bibinfo{author}{\bibfnamefont{Y.~B.} \bibnamefont{Kim}},
  \bibinfo{note}{cond-mat/0503241 (unpublished)}.

\bibitem[{\citenamefont{Lee and Lee}()}]{Lee}
\bibinfo{author}{\bibfnamefont{S.-S.} \bibnamefont{Lee}} \bibnamefont{and}
  \bibinfo{author}{\bibfnamefont{P.~A.} \bibnamefont{Lee}},
  \bibinfo{note}{cond-mat/0502139 (unpublished)}.

\bibitem[{\citenamefont{Motrunich}()}]{Motrunich}
\bibinfo{author}{\bibfnamefont{O.~I.} \bibnamefont{Motrunich}},
  \bibinfo{note}{cond-mat/0412556 (unpublished)}.

\bibitem[{\citenamefont{Fisher and Lee}(1989)}]{duality}
\bibinfo{author}{\bibfnamefont{M.~P.~A.} \bibnamefont{Fisher}}
  \bibnamefont{and} \bibinfo{author}{\bibfnamefont{D.~H.} \bibnamefont{Lee}},
  \bibinfo{journal}{Phys.\ Rev.\ B} \textbf{\bibinfo{volume}{39}},
  \bibinfo{pages}{2756} (\bibinfo{year}{1989}).

\bibitem[{\citenamefont{Fradkin}(1989)}]{JordanWigner}
\bibinfo{author}{\bibfnamefont{E.}~\bibnamefont{Fradkin}},
  \bibinfo{journal}{Phys.\ Rev.\ Lett.} \textbf{\bibinfo{volume}{63}},
  \bibinfo{pages}{322} (\bibinfo{year}{1989}).

\bibitem[{\citenamefont{Heinonen}(1998)}]{Heinonen}
\bibinfo{editor}{\bibfnamefont{O.}~\bibnamefont{Heinonen}}, ed.,
  \emph{\bibinfo{title}{Composite Fermions: A Unified View of the Quantum Hall
  Regime}} (\bibinfo{publisher}{World Scientific, Singapore},
  \bibinfo{year}{1998}).

\bibitem[{\citenamefont{Lopez et~al.}(1994)\citenamefont{Lopez, Rojo, and
  Fradkin}}]{Lopez}
\bibinfo{author}{\bibfnamefont{A.}~\bibnamefont{Lopez}},
  \bibinfo{author}{\bibfnamefont{A.~G.} \bibnamefont{Rojo}}, \bibnamefont{and}
  \bibinfo{author}{\bibfnamefont{E.}~\bibnamefont{Fradkin}},
  \bibinfo{journal}{Phys.\ Rev.\ B} \textbf{\bibinfo{volume}{49}},
  \bibinfo{pages}{15139} (\bibinfo{year}{1994}).

\bibitem[{\citenamefont{Hands et~al.}(2004)\citenamefont{Hands, Kogut,
  Scorzato, and Strouthos}}]{Kogut}
\bibinfo{author}{\bibfnamefont{S.~J.} \bibnamefont{Hands}},
  \bibinfo{author}{\bibfnamefont{J.~B.} \bibnamefont{Kogut}},
  \bibinfo{author}{\bibfnamefont{L.}~\bibnamefont{Scorzato}}, \bibnamefont{and}
  \bibinfo{author}{\bibfnamefont{C.~G.} \bibnamefont{Strouthos}},
  \bibinfo{journal}{Phys.\ Rev.\ B} \textbf{\bibinfo{volume}{70}},
  \bibinfo{pages}{104501} (\bibinfo{year}{2004}).

\bibitem[{\citenamefont{Pasquier and Haldane}(1998)}]{PH}
\bibinfo{author}{\bibfnamefont{V.}~\bibnamefont{Pasquier}} \bibnamefont{and}
  \bibinfo{author}{\bibfnamefont{F.~D.~M.} \bibnamefont{Haldane}},
  \bibinfo{journal}{Nucl.\ Phys.\ B} \textbf{\bibinfo{volume}{516}},
  \bibinfo{pages}{719} (\bibinfo{year}{1998}).

\bibitem[{\citenamefont{Read}(1998)}]{Read}
\bibinfo{author}{\bibfnamefont{N.}~\bibnamefont{Read}},
  \bibinfo{journal}{Phys.\ Rev.\ B} \textbf{\bibinfo{volume}{58}},
  \bibinfo{pages}{16262} (\bibinfo{year}{1998}).

\bibitem[{\citenamefont{Halperin et~al.}(1993)\citenamefont{Halperin, Lee, and
  Read}}]{HLR}
\bibinfo{author}{\bibfnamefont{B.~I.} \bibnamefont{Halperin}},
  \bibinfo{author}{\bibfnamefont{P.~A.} \bibnamefont{Lee}}, \bibnamefont{and}
  \bibinfo{author}{\bibfnamefont{N.}~\bibnamefont{Read}},
  \bibinfo{journal}{Phys.\ Rev.\ B} \textbf{\bibinfo{volume}{47}},
  \bibinfo{pages}{7312} (\bibinfo{year}{1993}).

\bibitem[{\citenamefont{Murthy and Shankar}(2003)}]{MS}
\bibinfo{author}{\bibfnamefont{G.}~\bibnamefont{Murthy}} \bibnamefont{and}
  \bibinfo{author}{\bibfnamefont{R.}~\bibnamefont{Shankar}},
  \bibinfo{journal}{Rev.\ Mod.\ Phys.} \textbf{\bibinfo{volume}{75}},
  \bibinfo{pages}{1101} (\bibinfo{year}{2003}).

\bibitem[{\citenamefont{Lee}(1998)}]{DHL}
\bibinfo{author}{\bibfnamefont{D.-H.} \bibnamefont{Lee}},
  \bibinfo{journal}{Phys.\ Rev.\ Lett.} \textbf{\bibinfo{volume}{80}},
  \bibinfo{pages}{4745} (\bibinfo{year}{1998}).

\bibitem[{\citenamefont{Feigelman et~al.}(1993)\citenamefont{Feigelman,
  Geshkenbein, Ioffe, and Larkin}}]{Feigelman}
\bibinfo{author}{\bibfnamefont{M.~V.} \bibnamefont{Feigelman}},
  \bibinfo{author}{\bibfnamefont{V.~B.} \bibnamefont{Geshkenbein}},
  \bibinfo{author}{\bibfnamefont{L.~B.} \bibnamefont{Ioffe}}, \bibnamefont{and}
  \bibinfo{author}{\bibfnamefont{A.~I.} \bibnamefont{Larkin}},
  \bibinfo{journal}{Phys. Rev. B} \textbf{\bibinfo{volume}{48}},
  \bibinfo{pages}{16641} (\bibinfo{year}{1993}).

\bibitem[{\citenamefont{Kawamura}(1998)}]{KawamuraRev}
\bibinfo{author}{\bibfnamefont{H.}~\bibnamefont{Kawamura}},
  \bibinfo{journal}{J.\ Phys.: Condens.\ Matter} \textbf{\bibinfo{volume}{10}},
  \bibinfo{pages}{4707} (\bibinfo{year}{1998}).

\bibitem[{\citenamefont{Pelissetto and Vicari}(2002)}]{Vicari}
\bibinfo{author}{\bibfnamefont{A.}~\bibnamefont{Pelissetto}} \bibnamefont{and}
  \bibinfo{author}{\bibfnamefont{E.}~\bibnamefont{Vicari}},
  \bibinfo{journal}{Phys.\ Rep.} \textbf{\bibinfo{volume}{368}},
  \bibinfo{pages}{549} (\bibinfo{year}{2002}).

\bibitem[{\citenamefont{Delamotte et~al.}(2004)\citenamefont{Delamotte,
  Mouhanna, and Tissier}}]{NonpertRG}
\bibinfo{author}{\bibfnamefont{B.}~\bibnamefont{Delamotte}},
  \bibinfo{author}{\bibfnamefont{D.}~\bibnamefont{Mouhanna}}, \bibnamefont{and}
  \bibinfo{author}{\bibfnamefont{M.}~\bibnamefont{Tissier}},
  \bibinfo{journal}{Phys.\ Rev.\ B} \textbf{\bibinfo{volume}{69}},
  \bibinfo{pages}{134413} (\bibinfo{year}{2004}).

\bibitem[{\citenamefont{Calabrese et~al.}(2004)\citenamefont{Calabrese,
  Parruccini, Pelissetto, and Vicari}}]{Calabrese}
\bibinfo{author}{\bibfnamefont{P.}~\bibnamefont{Calabrese}},
  \bibinfo{author}{\bibfnamefont{P.}~\bibnamefont{Parruccini}},
  \bibinfo{author}{\bibfnamefont{A.}~\bibnamefont{Pelissetto}},
  \bibnamefont{and} \bibinfo{author}{\bibfnamefont{E.}~\bibnamefont{Vicari}},
  \bibinfo{journal}{Phys. Rev. B} \textbf{\bibinfo{volume}{70}},
  \bibinfo{pages}{174439} (\bibinfo{year}{2004}).

\bibitem[{\citenamefont{Sak}(1974)}]{Sak}
\bibinfo{author}{\bibfnamefont{J.}~\bibnamefont{Sak}}, \bibinfo{journal}{Phys.\
  Rev.\ B} \textbf{\bibinfo{volume}{10}}, \bibinfo{pages}{3957}
  (\bibinfo{year}{1974}).

\bibitem[{\citenamefont{Kawamura}(1988)}]{KawamuraRG}
\bibinfo{author}{\bibfnamefont{H.}~\bibnamefont{Kawamura}},
  \bibinfo{journal}{Phys.\ Rev.\ B} \textbf{\bibinfo{volume}{38}},
  \bibinfo{pages}{4916} (\bibinfo{year}{1988}).

\bibitem[{\citenamefont{Kawamura}(1992)}]{KawamuraNum}
\bibinfo{author}{\bibfnamefont{H.}~\bibnamefont{Kawamura}},
  \bibinfo{journal}{J.\ Phys.\ Soc.\ Jpn.} \textbf{\bibinfo{volume}{61}},
  \bibinfo{pages}{1299} (\bibinfo{year}{1992}).

\bibitem[{\citenamefont{Plumer and Mailhot}(1994)}]{tricritical}
\bibinfo{author}{\bibfnamefont{M.~L.} \bibnamefont{Plumer}} \bibnamefont{and}
  \bibinfo{author}{\bibfnamefont{A.}~\bibnamefont{Mailhot}},
  \bibinfo{journal}{Phys.\ Rev.\ B} \textbf{\bibinfo{volume}{50}},
  \bibinfo{pages}{16113} (\bibinfo{year}{1994}).

\bibitem[{\citenamefont{Boubcheur et~al.}(1996)\citenamefont{Boubcheur, Loison,
  and Diep}}]{PhaseDiagram}
\bibinfo{author}{\bibfnamefont{E.~H.} \bibnamefont{Boubcheur}},
  \bibinfo{author}{\bibfnamefont{D.}~\bibnamefont{Loison}}, \bibnamefont{and}
  \bibinfo{author}{\bibfnamefont{H.~T.} \bibnamefont{Diep}},
  \bibinfo{journal}{Phys.\ Rev.\ B} \textbf{\bibinfo{volume}{54}},
  \bibinfo{pages}{4165} (\bibinfo{year}{1996}).

\bibitem[{\citenamefont{Zukovic et~al.}(2002)\citenamefont{Zukovic, Idogaki,
  and Takeda}}]{biquadratic}
\bibinfo{author}{\bibfnamefont{M.}~\bibnamefont{Zukovic}},
  \bibinfo{author}{\bibfnamefont{T.}~\bibnamefont{Idogaki}}, \bibnamefont{and}
  \bibinfo{author}{\bibfnamefont{K.}~\bibnamefont{Takeda}},
  \bibinfo{journal}{Phys.\ Rev.\ B} \textbf{\bibinfo{volume}{65}},
  \bibinfo{pages}{144410} (\bibinfo{year}{2002}).

\bibitem[{\citenamefont{Itakura}(2003)}]{Itakura}
\bibinfo{author}{\bibfnamefont{M.}~\bibnamefont{Itakura}},
  \bibinfo{journal}{J.\ Phys.\ Soc.\ Jpn.} \textbf{\bibinfo{volume}{72}},
  \bibinfo{pages}{74} (\bibinfo{year}{2003}).

\bibitem[{\citenamefont{Peles et~al.}(2004)\citenamefont{Peles, Southern,
  Delamotte, Mouhanna, and Tissier}}]{Peles}
\bibinfo{author}{\bibfnamefont{A.}~\bibnamefont{Peles}},
  \bibinfo{author}{\bibfnamefont{B.~W.} \bibnamefont{Southern}},
  \bibinfo{author}{\bibfnamefont{B.}~\bibnamefont{Delamotte}},
  \bibinfo{author}{\bibfnamefont{D.}~\bibnamefont{Mouhanna}}, \bibnamefont{and}
  \bibinfo{author}{\bibfnamefont{M.}~\bibnamefont{Tissier}},
  \bibinfo{journal}{Phys. Rev. B} \textbf{\bibinfo{volume}{69}},
  \bibinfo{pages}{220408} (\bibinfo{year}{2004}).

\bibitem[{\citenamefont{Loison and Schotte}(1998)}]{modifiedXY}
\bibinfo{author}{\bibfnamefont{D.}~\bibnamefont{Loison}} \bibnamefont{and}
  \bibinfo{author}{\bibfnamefont{K.~D.} \bibnamefont{Schotte}},
  \bibinfo{journal}{Eur.\ Phys.\ J.\ B} \textbf{\bibinfo{volume}{5}},
  \bibinfo{pages}{735} (\bibinfo{year}{1998}).

\bibitem[{\citenamefont{Pelissetto et~al.}(2001)\citenamefont{Pelissetto,
  Rossi, and Vicari}}]{6loop1}
\bibinfo{author}{\bibfnamefont{A.}~\bibnamefont{Pelissetto}},
  \bibinfo{author}{\bibfnamefont{P.}~\bibnamefont{Rossi}}, \bibnamefont{and}
  \bibinfo{author}{\bibfnamefont{E.}~\bibnamefont{Vicari}},
  \bibinfo{journal}{Phys.\ Rev.\ B} \textbf{\bibinfo{volume}{63}},
  \bibinfo{pages}{140414(R)} (\bibinfo{year}{2001}).

\bibitem[{\citenamefont{Calabrese et~al.}(2002)\citenamefont{Calabrese,
  Parruccini, and Sokolov}}]{6loop2}
\bibinfo{author}{\bibfnamefont{P.}~\bibnamefont{Calabrese}},
  \bibinfo{author}{\bibfnamefont{P.}~\bibnamefont{Parruccini}},
  \bibnamefont{and} \bibinfo{author}{\bibfnamefont{A.~I.}
  \bibnamefont{Sokolov}}, \bibinfo{journal}{Phys.\ Rev.\ B}
  \textbf{\bibinfo{volume}{66}}, \bibinfo{pages}{180403(R)}
  (\bibinfo{year}{2002}).

\bibitem[{\citenamefont{Lannert et~al.}(2001)\citenamefont{Lannert, Fisher, and
  Senthil}}]{Lannert}
\bibinfo{author}{\bibfnamefont{C.}~\bibnamefont{Lannert}},
  \bibinfo{author}{\bibfnamefont{M.~P.~A.} \bibnamefont{Fisher}},
  \bibnamefont{and} \bibinfo{author}{\bibfnamefont{T.}~\bibnamefont{Senthil}},
  \bibinfo{journal}{Phys.~Rev.~B} \textbf{\bibinfo{volume}{63}},
  \bibinfo{pages}{134510} (\bibinfo{year}{2001}).

\bibitem[{\citenamefont{Appelquist et~al.}(1986)\citenamefont{Appelquist,
  Bowick, Karabali, and Wijewardhana}}]{Appelquist}
\bibinfo{author}{\bibfnamefont{T.~W.} \bibnamefont{Appelquist}},
  \bibinfo{author}{\bibfnamefont{M.}~\bibnamefont{Bowick}},
  \bibinfo{author}{\bibfnamefont{D.}~\bibnamefont{Karabali}}, \bibnamefont{and}
  \bibinfo{author}{\bibfnamefont{L.~C.~R.} \bibnamefont{Wijewardhana}},
  \bibinfo{journal}{Phys.\ Rev.\ D} \textbf{\bibinfo{volume}{33}},
  \bibinfo{pages}{3704} (\bibinfo{year}{1986}).

\bibitem[{\citenamefont{Hermele et~al.}()\citenamefont{Hermele, Senthil, and
  Fisher}}]{Hermele}
\bibinfo{author}{\bibfnamefont{M.}~\bibnamefont{Hermele}},
  \bibinfo{author}{\bibfnamefont{T.}~\bibnamefont{Senthil}}, \bibnamefont{and}
  \bibinfo{author}{\bibfnamefont{M.~P.~A.} \bibnamefont{Fisher}},
  \bibinfo{note}{cond-mat/0502215 (unpublished)}.

\bibitem[{\citenamefont{Rantner and Wen}(2002)}]{Rantner}
\bibinfo{author}{\bibfnamefont{W.}~\bibnamefont{Rantner}} \bibnamefont{and}
  \bibinfo{author}{\bibfnamefont{X.-G.} \bibnamefont{Wen}},
  \bibinfo{journal}{Phys.\ Rev.\ B} \textbf{\bibinfo{volume}{66}},
  \bibinfo{pages}{144501} (\bibinfo{year}{2002}).

\bibitem[{\citenamefont{Franz et~al.}(2003)\citenamefont{Franz, Pereg-Barnea,
  Sheehy, and Tesanovic}}]{Franz}
\bibinfo{author}{\bibfnamefont{M.}~\bibnamefont{Franz}},
  \bibinfo{author}{\bibfnamefont{T.}~\bibnamefont{Pereg-Barnea}},
  \bibinfo{author}{\bibfnamefont{D.~E.} \bibnamefont{Sheehy}},
  \bibnamefont{and}
  \bibinfo{author}{\bibfnamefont{Z.}~\bibnamefont{Tesanovic}},
  \bibinfo{journal}{Phys.\ Rev.\ B} \textbf{\bibinfo{volume}{68}},
  \bibinfo{pages}{024508} (\bibinfo{year}{2003}).

\bibitem[{\citenamefont{Kaveh and Herbut}(unpublished)}]{Herbut}
\bibinfo{author}{\bibfnamefont{K.}~\bibnamefont{Kaveh}} \bibnamefont{and}
  \bibinfo{author}{\bibfnamefont{I.~F.} \bibnamefont{Herbut}},
  \bibinfo{journal}{cond-mat/0411594}  (\bibinfo{year}{unpublished}).

\bibitem[{\citenamefont{Franz et~al.}(2002)\citenamefont{Franz, Tesanovic, and
  Vafek}}]{Vafek}
\bibinfo{author}{\bibfnamefont{M.}~\bibnamefont{Franz}},
  \bibinfo{author}{\bibfnamefont{Z.}~\bibnamefont{Tesanovic}},
  \bibnamefont{and} \bibinfo{author}{\bibfnamefont{O.}~\bibnamefont{Vafek}},
  \bibinfo{journal}{Phys.\ Rev.\ B} \textbf{\bibinfo{volume}{66}},
  \bibinfo{pages}{054535} (\bibinfo{year}{2002}).

\bibitem[{\citenamefont{Borokhov
  et~al.}(2002{\natexlab{a}})\citenamefont{Borokhov, Kapustin, and Wu}}]{BKW}
\bibinfo{author}{\bibfnamefont{V.}~\bibnamefont{Borokhov}},
  \bibinfo{author}{\bibfnamefont{A.}~\bibnamefont{Kapustin}}, \bibnamefont{and}
  \bibinfo{author}{\bibfnamefont{X.}~\bibnamefont{Wu}}, \bibinfo{journal}{J.\
  High Energy Phys.} \textbf{\bibinfo{volume}{11}}, \bibinfo{pages}{049}
  (\bibinfo{year}{2002}{\natexlab{a}}).

\bibitem[{\citenamefont{Jackiw}(1984)}]{ZeroModes}
\bibinfo{author}{\bibfnamefont{R.}~\bibnamefont{Jackiw}},
  \bibinfo{journal}{Phys.\ Rev.\ D} \textbf{\bibinfo{volume}{29}},
  \bibinfo{pages}{2375} (\bibinfo{year}{1984}).

\bibitem[{\citenamefont{Lee and Fisher}(1989)}]{FQHvort1}
\bibinfo{author}{\bibfnamefont{D.}~\bibnamefont{Lee}} \bibnamefont{and}
  \bibinfo{author}{\bibfnamefont{M.~P.~A.} \bibnamefont{Fisher}},
  \bibinfo{journal}{Phys.\ Rev.\ Lett.} \textbf{\bibinfo{volume}{63}},
  \bibinfo{pages}{903} (\bibinfo{year}{1989}).

\bibitem[{\citenamefont{Lee and Fisher}(1991)}]{FQHvort2}
\bibinfo{author}{\bibfnamefont{D.}~\bibnamefont{Lee}} \bibnamefont{and}
  \bibinfo{author}{\bibfnamefont{M.~P.~A.} \bibnamefont{Fisher}},
  \bibinfo{journal}{Int.\ J.\ Mod.\ Phys.\ B} \textbf{\bibinfo{volume}{5}},
  \bibinfo{pages}{2675} (\bibinfo{year}{1991}).

\bibitem[{\citenamefont{Wen and Zee}(1992)}]{FQHvort3}
\bibinfo{author}{\bibfnamefont{X.~G.} \bibnamefont{Wen}} \bibnamefont{and}
  \bibinfo{author}{\bibfnamefont{A.}~\bibnamefont{Zee}},
  \bibinfo{journal}{Phys.\ Rev.\ B} \textbf{\bibinfo{volume}{46}},
  \bibinfo{pages}{2290} (\bibinfo{year}{1992}).

\bibitem[{\citenamefont{Aharony et~al.}(1997)\citenamefont{Aharony, Hanany,
  Intriligator, Seiberg, and Strassler}}]{Aharony}
\bibinfo{author}{\bibfnamefont{O.}~\bibnamefont{Aharony}},
  \bibinfo{author}{\bibfnamefont{A.}~\bibnamefont{Hanany}},
  \bibinfo{author}{\bibfnamefont{K.~A.} \bibnamefont{Intriligator}},
  \bibinfo{author}{\bibfnamefont{N.}~\bibnamefont{Seiberg}}, \bibnamefont{and}
  \bibinfo{author}{\bibfnamefont{M.~J.} \bibnamefont{Strassler}},
  \bibinfo{journal}{Nucl. Phys. B} \textbf{\bibinfo{volume}{499}},
  \bibinfo{pages}{67} (\bibinfo{year}{1997}).

\bibitem[{\citenamefont{Borokhov
  et~al.}(2002{\natexlab{b}})\citenamefont{Borokhov, Kapustin, and Wu}}]{BKW2}
\bibinfo{author}{\bibfnamefont{V.}~\bibnamefont{Borokhov}},
  \bibinfo{author}{\bibfnamefont{A.}~\bibnamefont{Kapustin}}, \bibnamefont{and}
  \bibinfo{author}{\bibfnamefont{X.}~\bibnamefont{Wu}}, \bibinfo{journal}{J.
  High Energy Phys.} \textbf{\bibinfo{volume}{12}}, \bibinfo{pages}{044}
  (\bibinfo{year}{2002}{\natexlab{b}}).

\bibitem[{\citenamefont{Kim et~al.}(1997)\citenamefont{Kim, Lee, and
  Lee}}]{DHKim}
\bibinfo{author}{\bibfnamefont{D.~H.} \bibnamefont{Kim}},
  \bibinfo{author}{\bibfnamefont{D.~K.~K.} \bibnamefont{Lee}},
  \bibnamefont{and} \bibinfo{author}{\bibfnamefont{P.~A.} \bibnamefont{Lee}},
  \bibinfo{journal}{Phys. Rev. B} \textbf{\bibinfo{volume}{55}},
  \bibinfo{pages}{591} (\bibinfo{year}{1997}).

\end{thebibliography}

\end{document}